\documentclass[12 pt]{article}

\usepackage{graphicx,latexsym,amsmath,amsfonts}
\usepackage{amssymb,amsmath}

\usepackage{amsthm}
\usepackage{enumerate}
\usepackage{epsfig}
\usepackage{verbatim}

\usepackage{epsfig}

\newcommand{\fracs}[2]{{ \textstyle \frac{#1}{#2} }}
\newtheorem{theorem}{Theorem}[section]

\newtheorem{assp}[theorem]{Assumption}

\newcommand{\fp}{{f}}
\newcommand{\fm}{{a}}

\newcommand{\PP}{\mathbb{P}}
\newcommand{\RR}{\mathbb{R}}
\newcommand{\NN}{\mathbb{N}}

\newtheorem{lemma}[theorem]{Lemma}

\newcommand{\eps}{\varepsilon}

\newcommand{\Levy}{L{\'e}vy }

\newcommand{\hP}{\widehat{P}}

\newcommand{\EE}{\mathbb{E}}
\newcommand{\VV}{\mathbb{V}}

\def \Levy {{L{\'e}vy }}

\def \D {\Delta}

\def \X {\overline{X}}

\def \eps {\epsilon}

\def \lev  {\left\|} \def\rev{\right\|}

\def \Eps {\varepsilon}

\newcommand{\halfs}{{\scriptstyle \frac{1}{2}}}

\def\a{\alpha} 
\def\e{\varepsilon}   \def\o{\theta}
 \def\k{\kappa} \def\l{\lambda}  
   \def\s{\sigma}

\def\be{\beta}

\DeclareMathOperator{\bO}{{             {\cal O}}}
\DeclareMathOperator{\bo}{{\scriptscriptstyle {\cal O}}}

\def \hP {{\widehat{P}}}

\def\1{\mathbf{1}}
\newcommand{\tX}{ {\tilde X} }

\begin{document}

\title{Multilevel Monte Carlo methods\\ for applications in finance}
\author{Mike Giles and Lukasz Szpruch\\[0.1in]
        Oxford-Man Institute of Quantitative Finance\\
        and Mathematical Institute, University of Oxford}

\maketitle

\begin{abstract}
Since Giles introduced the multilevel Monte Carlo path simulation 
method \cite{giles08}, there has been rapid development of the technique for
a variety of applications in computational finance.  This paper surveys
the progress so far, highlights the key features in achieving a high
rate of multilevel variance convergence, and suggests directions for 
future research.
\end{abstract}

\section{Introduction}

In 2001, Heinrich \cite{heinrich01},  developed a multilevel Monte Carlo method for
parametric integration, in which one is interested in estimating
the value of
$\EE[f(x,\lambda)]$ where $x$ is a finite-dimensional random
variable and $\lambda$ is a parameter.  In the simplest case in
which $\lambda$ is a real variable in the range $[0,1]$, having
estimated the value of $\EE[f(x,0)]$ and $\EE[f(x,1)]$, one
can use $\frac{1}{2}(f(x,0)+f(x,1))$ as a control variate when
estimating the value of $\EE[f(x,\frac{1}{2})]$, since the
variance of
$
f(x,\frac{1}{2}) - \frac{1}{2}(f(x,0)+f(x,1))
$
will usually be less than the variance of $f(x,\frac{1}{2})$.
This approach can then be applied recursively for other intermediate
values of $\lambda$, yielding large savings if $f(x,\lambda)$ is
sufficiently smooth with respect to $\lambda$.

Giles' multilevel Monte Carlo path simulation \cite{giles08} is
both similar and different.  There is no parametric integration,
and the random variable is infinite-dimensional, corresponding to
a Brownian path in the original paper.  However, the control variate
viewpoint is very similar. A coarse path simulations is used as a
control variate for a more refined fine path simulation, but since
the exact expectation for the coarse path is not known, this is in
turn estimated recursively using even coarser path simulation as
control variates.  The coarsest path in the multilevel hierarchy
may have only one timestep for the entire interval of interest.

A similar two-level strategy was developed slightly earlier by
Kebaier \cite{kebaier05}, and a similar multi-level approach was under
development at the same time by Speight
\cite{speight09,speight10}.

In this review article, we start by introducing the central ideas
in multilevel Monte Carlo simulation, and the key theorem from
\cite{giles08} which gives the greatly improved computational cost
if a number of conditions are satisfied.  The challenge then is to
construct numerical methods which satisfy these conditions, and we
consider this for a range of computational finance applications.

\section{Multilevel Monte Carlo}
\subsection{Monte Carlo}

Monte Carlo simulation has become an essential tool in the pricing of derivatives security and in risk management. 
In the abstract setting, our goal is to numerically approximate the expected value  $\EE [Y]$, where $Y = P (X) $
is a functional of a random variable $X$. In most financial applications we are not able to sample $X$ directly and 
hence, in order to perform Monte Carlo simulations we approximate $X$ with $X_{\D t}$ such that  
$\EE [P(X_{\D t}) ] \rightarrow \EE [ P(X) ] $, when $\D t \rightarrow 0$.    
Using $X_{\D t}$ to compute $N$ approximation samples produces the standard Monte Carlo estimate 
\[
 \hat{Y} = \frac{1}{N}\sum_{i=1}^{N} P(X_{\D t}^{i}),
\]
where $X_{\D t}^{i}$ is the numerical approximation to $X$ on the $i$th sample path  
and $N$ is the number of independent simulations of $X$.  By standard Monte Carlo results
$\hat{Y} \rightarrow \EE [Y]$, when $\D t \rightarrow 0$ and $N \rightarrow \infty$. 
In practice we perform Monte Carlo simulation with given $\D t>0$ and finite $N$ producing an error 
to the approximation of  $\EE [Y]$. Here we are interested in the mean square error that is 
\[
MSE \equiv \EE \left[  (  \hat{Y} - \EE [Y] )^{2}   \right]
\]
Our goal in the design of the Monte Carlo algorithm is to estimate $Y$ with accuracy root-mean-square error $\e$ ($MSE \le \e^{2}$), as efficiently as possible. That is to minimize the computational complexity required to achieve 
the desired mean square error. For standard Monte Carlo simulations the mean square error can be expressed as
\begin{equation*}
 \begin{split}
  \EE \left[  (  \hat{Y} - \EE [Y] )^{2}   \right] 
= & \EE \left[  (  \hat{Y} - \EE [\hat{Y}] + \EE[\hat{Y}]- \EE [Y] )^{2}   \right]\\
= & \underbrace{ \EE \left[  (  \hat{Y} - \EE [\hat{Y}] )^{2} \right] }_{\text{Monte Carlo variance }} 
+  \underbrace{\left(  \EE[\hat{Y}]- \EE [Y] )^{2}   \right)}_{\text{bias of the approximation}}.
 \end{split}
\end{equation*}
The Monte Carlo variance is proportional to $\frac{1}{N}$
\[
 \VV[\hat{Y}] = \frac{1}{N^{2}}\VV\left[ \sum_{i=1}^{N} P(X_{\D t}^{i}) \right] = \frac{1}{N} \VV[P(X_{\D t})]. 
\]
 For both Euler-Maruyama and Milstein approximation   $| \EE[\hat{Y}]- \EE [Y] |=\bO(\D t)$, typically. Hence 
the mean square error for standard Monte Carlo is given by
\[
 \EE \left[  (  \hat{Y} - \EE [Y] )^{2}   \right]  = \bO(\frac{1}{N}) + \bO(\D t^{2}). 
\]
To ensure the root-mean-square error is proportional to $\e$, we must have $MSE = \bO(\e^{2})$ and therefore
$1/N=\bO(\e^{2})$ and $\D t^{2}=\bO(\e^{2})$, which means $N=\bO(\e^{-2})$ and $\D t=\bO(\e)$. The computational cost 
of standard Monte Carlo is proportional to the number of paths $N$  multiplied by the cost of generating a path, that is 
the number of timesteps in each sample path. Therefore, the cost is $C=\bO(\e^{-3})$. In the next section we will show that using MLMC we can reduce the complexity of achieving root mean square error $\e$ to $\bO(\e^{-2})$.  
 
\subsection{Multilevel Monte Carlo Theorem}

In its most general form, multilevel Monte Carlo (MLMC) simulation uses a number of 
levels of resolution, $\ell=0,1,\ldots,L$, with $\ell=0$ being the coarsest, 
and  $\ell=L$ being the finest.  In the context of a SDE simulation, level $0$ 
may have just one timestep for the whole time interval $[0,T]$, whereas level
$L$ might have $2^L$ uniform timesteps $\D t_{L} = 2^{-L}T$.

%If $P=P((x(t))_{t\in [0, T]})$ 
If $P$ 
denotes the payoff (or other output functional of interest), and 
%$P_{\ell}=P((X_{t_{k}})_{0 \le k \le \ell})$ 
$P_{\ell}$ 
denotes its approximation on level $l$, then 
the expected value $\EE[P_L]$ on the finest level is equal to the expected value $\EE[P_0]$
on the coarsest level plus a sum of corrections which give the difference in 
expectation between simulations on successive levels,
\begin{equation} \label{eq:MLMC}
 \EE [P_{L}] = \EE [P_{0}] + \sum_{\ell=1}^{L}\EE [P_{\ell}-P_{\ell-1}].
\end{equation}
The idea behind MLMC is to independently estimate each of the expectations 
on the right-hand side of \eqref{eq:MLMC} in a way which minimises the overall 
variance for a given computational cost. Let $Y_0$ be an estimator for 
$\EE [P_{0}]$ using $N_{0}$  samples, and let $Y_\ell$, $\ell>0$, be an 
estimator for $\EE [P_{\ell} - P_{\ell-1}]$ using $N_{\ell}$ samples. The simplest 
estimator is a mean of $N_{\ell}$ independent samples, which for $\ell>0$ is
\begin{equation} \label{eq:est}
 Y_{\ell} = N_{\ell}^{-1}\sum_{i=1}^{N_{\ell}} (P^{i}_{\ell}-P^{i}_{\ell-1}).
\end{equation}
The key point here is that $P^{i}_{\ell}\!-\!P^{i}_{\ell-1}$ should come 
from two discrete approximations for the same underlying stochastic sample (see \cite{pages07}),
so that on finer levels of resolution the difference is small (due to strong
convergence) and so its variance is also small. Hence very few samples will be 
required on finer levels to accurately estimate the expected value.

The combined MLMC estimator $\hat{Y}$ is
\[
 \hat{Y} = \sum_{\ell = 0}^{L} Y_{\ell}.
\]
We can observe that
\[
 \EE [Y_{\ell}] = N_{\ell}^{-1}\sum_{i=1}^{N_{\ell}} \EE [P^{i}_{\ell}-P^{i}_{\ell-1}] = \EE [P^{i}_{\ell}-P^{i}_{\ell-1}],
\]
and
\[
 \EE [\hat{Y}] =  \sum_{\ell = 0}^{L} \EE [Y_{\ell}]  = \EE [P_{0}] + \sum_{\ell=1}^{L}\EE [P_{\ell}-P_{\ell-1}]= \EE [P_{L}].
\]
Although we are using different levels with different discretization errors to estimate $\EE[P]$, the final 
accuracy depends on the accuracy of the finest level $L$.

Here we recall the Theorem from \cite{giles08} (which is a slight generalisation of
the original theorem in \cite{giles08}) which gives the complexity of MLMC estimation.

\begin{theorem} \label{th:complexity}
Let $P$ denote a functional of the solution of a stochastic differential 
equation, and let $P_\ell$ denote the corresponding level $\ell$ numerical 
approximation.  If there exist independent estimators $Y_\ell$ based on 
$N_\ell$ Monte Carlo samples, and positive constants 
$\alpha, \beta, \gamma, c_1, c_2, c_3$ such that 
$\alpha\!\geq\!\fracs{1}{2}\,\min(\beta,\gamma)$ and
\begin{itemize}
\item[i)]
$\displaystyle
\left| \EE[P_\ell \!-\! P] \right| \leq c_1\, 2^{-\alpha\, \ell}
$
\item[ii)]
$\displaystyle
\EE[Y_\ell] = \left\{ \begin{array}{ll}
\EE[P_0], & \ell=0 \\[0.1in]
\EE[P_\ell \!-\! P_{\ell-1}], & \ell>0
\end{array}\right.
$
\item[iii)]
$\displaystyle
\VV[Y_\ell] \leq c_2\, N_\ell^{-1} 2^{-\beta\, \ell}
$
\item[iv)]
$\displaystyle
C_\ell \leq c_3\, N_\ell\, 2^{\gamma\, \ell},
$
where $C_\ell$ is the computational complexity of $Y_\ell$
\end{itemize}
then there exists a positive constant $c_4$ such that for any $\eps \!<\! e^{-1}$
there are values $L$ and $N_\ell$ for which the multilevel estimator
\[
Y = \sum_{\ell=0}^L Y_\ell,
\]
has a mean-square-error with bound
\[
MSE \equiv \EE\left[ \left(Y - \EE[P]\right)^2\right] < \eps^2
\]
with a computational complexity $C$ with bound
\[
C \leq \left\{\begin{array}{ll}
c_4\, \eps^{-2}              ,    & \beta>\gamma, \\[0.1in]
c_4\, \eps^{-2} (\log \eps)^2,    & \beta=\gamma, \\[0.1in]
c_4\, \eps^{-2-(\gamma\!-\!\beta)/\alpha}, & 0<\beta<\gamma.
\end{array}\right.
\]
\end{theorem}

%The original Theorem  in \cite{giles08} was proved with  $\gamma=1$. 
%Also $\gamma=2$ if we define time step to be $4^{-l}$, as it was suggested in  \cite{giles08}.

\subsection{Improved MLMC}

In the previous section we showed that the key step in MLMC analysis is 
the estimation of variance $\VV [  P^{i}_{\ell}-P^{i}_{\ell-1}] $. As it will become more clear in the next section,
this is related to the strong convergence results on approximations of SDEs, which differentiates MLMC 
from standard MC, where we only require a weak error bound for approximations of SDEs. 

We will demonstrate that in fact the classical strong convergence may not be necessary for a good MLMC variance. 
In \eqref{eq:est} we have used the same estimator for the payoff 
$P_{\ell}$ on every level $\ell$, and therefore \eqref{eq:MLMC} is a 
trivial identity due to the telescoping summation.  However, in 
\cite{giles08b} Giles demonstrated that it can be better to use different 
estimators for the finer and coarser of the two levels being 
considered, $P^{f}_{\ell}$ when level $\ell$ is the finer level, 
and $P^{c}_{\ell}$ when level $\ell$ is the coarser level.
In this case, we require that
\begin{equation} \label{con:MLMC}
\EE[P^{f}_{\ell}] = \EE[P^{c}_{\ell}] \quad \hbox{for } \ell=1,\ldots,L,
\end{equation}
so that 
\[
E [P^f_{L}] = \EE [P^f_{0}] + \sum_{\ell=1}^{L}\EE [P^{f}_{\ell}-P^{c}_{\ell-1}].
\]
The MLMC Theorem is still applicable to this modified estimator.  The 
advantage is that it gives the flexibility to construct approximations for 
which $P^{f}_{\ell}-P^{c}_{\ell-1}$ is much smaller than the original
$P_{\ell}-P_{\ell-1}$, giving a larger value for $\beta$, the rate of variance
convergence in condition {\it iii)} in the theorem. 
In the next sections we demonstrate how suitable choices of 
$P^{f}_{\ell}$ and $P^{c}_{\ell}$ can dramatically increase the convergence of
the variance of the MLMC estimator.  

The good choice of estimators, as we shall see, often follows from analysis of the problem under consideration 
from  the distributional point of view. We will demonstrate that methods that had been used previously to improve the weak order of convergence can also improve the order of convergence of the MLMC variance.   

\subsection{SDEs}

First, we consider a general class of $d$-dimensional SDEs driven by Brownian 
motion. These are the primary object of studies in mathematical finance. 
In subsequent sections we demonstrate extensions of MLMC beyond the Brownian setting. 

Let $(\Omega, {\mathcal{F}},\{{\mathcal{F}}_t\}_{t\geq 0}, \PP)$ 
be a complete probability space with a filtration
$\{{\mathcal{F}}_t\}_{t\geq 0}$ satisfying the usual conditions,
%that is to say, it is right continuous and increasing while
%${\mathcal{F}}_0$ contains all $\PP$-null sets.  
and let 
%$w(t)=(w_{1}(t),...,w_{m}(t))^{T}$ 
$w(t)$ 
be a $m$-dimensional 
Brownian motion defined on the probability space.  
We consider the 
numerical approximation of SDEs of the form
\begin{equation}   \label{eq:SDE}
{\rm d}x(t)=f(x(t))\ {\rm d}t+g(x(t))\ {\rm d}w(t),
\end{equation}
where $x(t)\in \RR^{d}$ for each $t\!\ge\! 0$,
$f\in C^2 ( \RR^{d}, \RR^{d})$,
$g\in C^2 (\RR^{d}, \RR^{d\times m})$,
and for simplicity we assume a fixed initial value 
$x_{0}\in \RR^{d}$. 
The most prominent example of SDEs in finance is a geometric Brownian motion
\[
 {\rm d}x(t) = \a x(t)\,{\rm d}t + \be x(t)\, {\rm d}w(t), 
\]
where $\a,\be>0$. Although, we can solve this equation explicitly it is still 
worthwhile to approximate its solution numerically in order to judge the performance 
of the numerical procedure we wish to apply to more complex problems.  
Another interesting example is the famous Heston stochastic volatility model 
\begin{equation} \label{eq:Heston}
\begin{cases}
{\rm d}s(t) = r s(t)\,{\rm d}t + s(t)\sqrt{v(t)}\,{\rm d} w_{1}(t) \\
{\rm d}v(t) = \k(\o-v(t))\,{\rm d}t + \s\sqrt{v(t)}\,{\rm d} w_{2}(t) \\
\,{\rm d}w_{1}\,{\rm d}w_{2} = \rho \,{\rm d}\,t,
\end{cases}
\end{equation}
where $r,\k,\o,\s>0$. In this case we do not know the explicit form of the solution 
and therefore numerical integration is essential in order to price certain 
financial derivatives using the Monte Carlo method. At this point we would like to point out 
that the Heston model \eqref{eq:Heston} does not satisfy standard conditions required for numerical 
approximations to converge. Nevertheless, in this paper we always assume that coefficients 
of SDEs \eqref{eq:SDE} are sufficiently smooth. We refer to \cite{kloeden2012convergence,mao2012strong,szpruch2011numerical}
for an overview of the methods that can be applied when the global Lipschitz condition does not hold. 
We also refer the reader to \cite{kloeden2011multilevel} for an application of MLMC to the SDEs with additive fractional noise. 

\subsection{Euler and Milstein discretisations}

The simplest approximation of SDEs \eqref{eq:SDE} is an Euler-Maruyama (EM) scheme. 
Given any step size $\Delta
t_{\ell}$,  we define the partition $\mathcal{P}_{\Delta
t_{\ell}}:=\{n\Delta t_{\ell} :n=0,1,2,...,2^{\ell}\}$ of the time interval $[0,T]$, $2^{\ell}\D t = T>0$. The EM approximation 
$X^{\ell}_n \approx x(n\,\D t_{\ell})$ has the form \cite{kp92}
\begin{equation} \label{eq:EM}
 X^{\ell}_{n+1} = X^{\ell}_{n} + f (X^{\ell}_{n})\, \D t_{\ell} + g(X^{\ell}_{n})\,\D w^{\ell}_{n+1}, 
\end{equation}
where $\Delta w^{\ell}_{n+1} = w((n+1)\D t_{\ell} ) - w(n \D t_{\ell})  $ and $X_{0}=x_{0}$.
Equation \eqref{eq:EM} is written in a vector form and its
$i^{th}$ component reads as
\begin{equation*}   
 X^{\ell}_{i,n+1} = X^{\ell}_{i,n} + f_i(X^{\ell}_n)\, \D t_{\ell} 
      + \sum_{j=1}^m   g_{ij}(X^{\ell}_n)\, \D w^{\ell}_{j,n+1}.      
\end{equation*}
In the classical Monte Carlo setting we are mainly interested in the weak approximation of SDEs \eqref{eq:SDE}. Given 
a smooth payoff $P: \RR^{d} \rightarrow \RR$ we say that $X^{\ell}_{2^{\ell}}$ converges to $x(T)$ in a weak sense with order $\a$
if
\[
 | \EE [ P(x(T )) ] - \EE [ P(X^{\ell}_{T}) ]  | = \bO(\D t_{\ell} ^{\a}).
\]
Rate $\a$ is required in condition $(i)$ of Theorem \ref{th:complexity}. 
However, for MLMC condition $(iii)$ of Theorem \ref{th:complexity} 
is crucial.% (Clearly if $(iii)$ holds with $\be$ then 
%$(i)$ holds with at least $\be$).
 We have 
\[
 \VV_{\ell} \equiv {\rm Var\,}( P_\ell \!-\! P_{\ell-1}  ) \le \EE \left[ ( P_\ell \!-\! P_{\ell-1} )^{2} \right], 
\]
and
\[
 \EE \left[ ( P_\ell \!-\! P_{\ell-1} )^{2} \right] \le 2 \, \EE \left[  ( P_\ell \!-\! P )^{2} \right] 
    + 2\, \EE  \left[ ( P \!-\! P_{\ell-1} )^{2} \right].
\]
For Lipschitz continuous payoffs, $( P(x) - P(y) )^2\le L \lev x-y \rev^2 $, we then have
\[
 \EE \left[  ( P_\ell \!-\! P )^{2} \right] \le L \, \EE \left[ \lev  x(T) \!-\! X^{\ell}_{T} \rev^{2}\right]. 
\]  
It is clear now, that in order to estimate the variance of the MLMC we need to examine strong convergence property. The classical strong convergence on the finite time interval 
$[0,T]$ is defined as
\[
\left( \EE \left[  \lev x(T)-X^{\ell}_{T} \rev^{p}\right] \right)^{1/p}   = \bO(\D t_{\ell}^\xi), \quad \text{for} \quad p\ge2. 
\]
For the EM scheme $\xi = 0.5 $. In order to deal with path dependent options
we often require measure the error in the supremum norm: 
 \begin{equation*}
 \left(\EE \bigl[\sup_{0\le n \le 2^{\ell}}  \lev x(n \D t_{\ell})-X^{\ell}_n \rev^{p}  \bigr] \right)^{1/p} 
= \bO (\D t^{\xi}) \quad \text{for} \quad p\ge2.
 \end{equation*}
 % For some path-dependent payoff it is useful to define piecewise linear interpolation of EM scheme, i.e.
% \[
%  X(n \D t +\o \D t ) = (1-\o)X_{n} + \o X_{n+1} \quad \text{for } 0\le  \o <1, \, k=0,1,2,...,N-1.   
% \]
% M\"uller-Gronbach established that
%  \begin{equation*}
%  \EE \bigl[\sup_{0\le t\le T} \lev x(t)-X(t) \rev^{p}  \bigr] = O( \mid \D t\log(\D t)\mid^{p/2})
%  \end{equation*}
% % the following Theorem.  
% % \begin{theorem}[]
% %  Under global Lipschitz condition on functions $f$ and $g$, for any $p\ge1$ and
% %  $T\ge0$, there exists a positive constant $C(p,T)$, independent of $\D t$ such that
% %  \begin{equation*}
% %  \EE \bigl[\sup_{0\le t\le T} \lev x(t)-X(t) \rev^{p}  \bigr]\le C(p,T) \mid \D t\log(\D t)\mid^{p/2}
% %  \end{equation*}
% % \end{theorem}
Even in the case of globally Lipschitz continuous payoff $P$, the  
EM does not achieve $\be = 2\xi >1$ which is optimal 
in Theorem \eqref{th:complexity}. In order to improve the convergence of the MLMC variance the  
Milstein approximation $X_n \approx x(n\,\D t_{\ell})$ is considered, with $i^{th}$ component of the form \cite{kp92}
\begin{equation}   \label{eq:Milstein}
\begin{split}
 X^{\ell}_{i,n+1} = & X^{\ell}_{i,n} + f_i(X^{\ell}_n)\, \D t_{\ell} 
      + \sum_{j=1}^m   g_{ij}(X^{\ell}_n)\, \D w^{\ell}_{j,n+1} \\
    &  + \sum_{j,k=1}^m h_{ijk}(X^{\ell}_n) 
\left(\D w^{\ell}_{j,n}\D w^{\ell}_{k,n} - \Omega_{jk}\, \D t_{\ell} - A^{\ell}_{jk,n} \right)
\end{split}
\end{equation}
where 
$\Omega$ is the correlation matrix for the driving Brownian paths,
and $A^{\ell}_{jk,n}$ is the \Levy area defined as
\[
A^{\ell}_{jk,n} = \int_{n\D t_{\ell}}^{(n+1)\D t_{\ell}} \left( \rule{0in}{0.14in}
\left(\rule{0in}{0.13in} w_j(t) \!-\! w_j(n\D t_{\ell}) \right) {\rm d} w_k(t) -
\left(\rule{0in}{0.13in} w_k(t) \!-\! w_k(n\D t_{\ell}) \right) {\rm d} w_j(t)
\right).
\]
% \begin{theorem}[Milstein, Wagner and Platen]
%  Under global Lipschitz condition on functions $f$ and $g$, for any $p\ge1$ and
%  $T\ge0$, there exists a positive constant $C(p,T)$, independent of $\D t$ such that
%  \begin{equation*}
%  \EE \bigl[\sup_{0\le t_{k}\le T} \lev x(t_{k})-X_{t_{k}} \rev^{p}  \bigr]\le C(p,T) \D t^{p}.
%  \end{equation*}
% \end{theorem}
The rate of strong convergence $\xi$ for the Milstein scheme is double the value we have for the EM scheme and therefore
the MLMC variance for Lipschitz payoffs converges twice as fast. However, this gain does not come without a price. 
There is no efficient method  to simulate \Levy areas, apart from  dimension 2 \cite{gl94,rw01,wiktorsson01}.
In some applications, the diffusion coefficient $g(x)$ satisfies a commutativity 
property which gives
\[
h_{ijk}(x) = h_{ikj}(x) \quad \hbox{for all} \quad i, j, k.
\]
In that case, because the \Levy areas are anti-symmetric 
(i.e. $A^{l}_{jk,n}=-A^{l}_{kj,n}$), it follows that 
$h_{ijk}(X^{\ell}_n)\, A^{l}_{jk,n} + h_{ikj}(X^{{\ell}}_n)\, A^{l}_{kj,n} = 0$
and therefore the terms involving the \Levy areas cancel and so it is not
necessary to simulate them.  
However, this only happens in special cases.
Clark \& Cameron \cite{cc80} proved for a particular SDE that it is 
impossible to achieve a better order of strong convergence than the 
Euler-Maruyama discretisation when using just the discrete increments 
of the underlying Brownian motion. The analysis was extended by 
M\"{u}ller-Gronbach \cite{muller02} to general SDEs. 
As a consequence if we use the standard MLMC method with the Milstein 
scheme without simulating the \Levy areas the complexity will remain 
the same as for Euler-Maruyama.  Nevertheless, Giles and Szpruch showed in \cite{giles2012antithetic}
that by constructing a suitable antithetic estimator one can neglect the 
\Levy areas and still obtain a multilevel correction estimator with a 
variance which decays at the same rate as the scalar Milstein estimator.
% In the special case of SDEs \eqref{eq:SDE} with $d=m=1$ Milstein scheme has a form
% \begin{equation} \label{eq:M1D}
% X_{n+1} = X_{n} + f (X_{n}) \D t + g(X_{n})\D w_{n+1} + g^{'}(X_{n})g(X_{n}) (   \D w_{n+1}^{2} - \D t  ), 
% \end{equation}
% where $g^{'}=\fracs{dg}{dx}$.

\subsection{MLMC algorithm }

Here we explain how to implement the Monte Carlo algorithm. Let us recall that the
MLMC estimator $Y$ is given by
\[
 \hat{Y} = \sum_{\ell = 0}^{L} Y_{\ell}.
\]
We aim to minimize the computational cost necessary to achieve desirable accuracy $\e$. As for standard Monte Carlo
we have 
\begin{equation*}
 \begin{split}
  \EE \left[  (  Y - \EE [P(X)] )^{2}   \right] 
= & \underbrace{ \EE \left[  (  Y - \EE [\hat{Y}] )^{2} \right] }_{\text{Monte Carlo variance }} 
+  \underbrace{\left(  \EE[P_{L}]- \EE [P(X)] )^{2}   \right)}_{\text{bias of the approximation}}.
 \end{split}
\end{equation*}
The variance is given by 
\[
 \VV[Y] = \sum_{\ell=0}^{L} \VV[Y_{\ell}] = \sum_{\ell=0}^{L} \frac{1}{N_{\ell}} \,V_{\ell},
\]
where $V_{\ell}  = \VV[P_{\ell}-P_{\ell-1}]$. To minimize the variance of $Y$ for fixed computational 
cost  $C= \sum_{\ell=0}^{L}N_{\ell}\D t_{\ell}^{-1}$, we can treat $N_{\ell}$ as continuous variable 
and use the Lagrange function to find the minimum of 
\[
 L = \sum_{\ell=0}^{L} \frac{1}{N_{\ell}} V_{\ell} + \lambda \left( \sum_{\ell=0}^{L}N_{\ell}\D t_{\ell}^{-1} - C\right). 
\]
First order conditions shows that $N_{\ell}= \lambda^{-\fracs{1}{2}}\sqrt{V_{\ell}\D t_{\ell}}$, therefore 
\[
 \VV[Y] = \sum_{\ell=0}^{L} \frac{V_{\ell}}{N_{\ell}} 
= \sum_{\ell=0}^{L} \frac{\sqrt{\lambda}}{\sqrt{V_{\ell}\D t_{\ell}}} V_{\ell}. 
\]
Since we want $\VV[Y] \le \fracs{\e^2}{2}$ we can show that
\[
 \lambda^{-\fracs{1}{2}}\ge 2 \e^{-2} \sum_{\ell=0}^{L}\sqrt{V_{\ell}/\D t_{\ell}}, 
\]
thus the optimal number of samples for level $\ell$ is
\begin{equation} \label{eq:optimalN}
 N_{\ell} = \left\lceil   2 \e^{-2} \sqrt{V_{\ell}\D t_{\ell}} \sum_{\ell=0}^{L}\sqrt{V_{\ell}/\D t_{\ell}}    \right\rceil.
\end{equation}
Assuming $\bO(\D t_{\ell})$ weak convergence, the bias of the overall method is 
equal $ c \D t_{L}=c\,T\,2^{-L}$. If we want the bias to be proportional to $\fracs{\e}{\sqrt{2}}$ we set
\[
 L_{\max} =\frac{\log{(\e/(c T \sqrt{2}))^{-1}}}{\log{2}}.
\]
From here we can calculate the overall complexity. 
We can now outline the algorithm
\begin{enumerate}
\item Begin with L=0;
\item Calculate the initial estimate of $V_{L}$ using 100 samples. 
\item Determine optimal $N_{\ell}$ using \eqref{eq:optimalN}.
\item Generate additional samples as needed for new $N_{\ell}$.
\item if $L< L_{\max}$ set $L := L+1$ and go to 2. 
\end{enumerate}

Most numerical tests suggests that $L_{\max}$ is not optimal and we can substantially improve MLMC 
by determining optimal $L$ by looking at bias. For more details see \cite{giles08}. 

\section{Pricing with MLMC}

A key application of MLMC is to compute the expected payoff of financial options. 
We have demonstrated that for globally Lipschitz European  payoffs, convergence of the MLMC variance is determined
by the strong rate of convergence of the corresponding numerical scheme. However, in many financial applications
payoffs are not smooth or are path-dependent. The aim of this section is to overview results on  mean square
convergence rates for Euler--Maruyama and Milstein approximations with
more complex payoffs. 
In the case of EM, the majority of payoffs encountered in practice have been analyzed in Giles et al. \cite{ghm09}. 
Extension of this analysis to the Milstein scheme is far from obvious. This is due to the fact that Milstein scheme gives 
an improved rate of convergence on the grid points, but this is insufficient for path dependent options. 
In many applications the behaviour of the numerical approximation between grid points is crucial. 
The analysis of Milstein scheme for complex payoffs was carried out in \cite{giles_ancient}.
To understand this problem better, we recall a few facts from the theory of strong convergence 
of numerical approximations. 
We can define a piecewise linear interpolation of a numerical approximation
 within the time interval  $[n \D t_{\ell}, (n+1) \D t_{\ell} )$ as
\begin{equation} \label{eq:pclint}
  X^{\ell}(t) = X^{\ell}_{n} + \l_{\ell} (X^{{\ell}}_{n+1} - X^{\ell}_{n} ),  \quad \text{for} \quad t\in [n \D t_{\ell}, (n+1) \D t_{\ell} )
\end{equation}
where $\l_{\ell} \equiv (t-n\D t_{\ell})/\D t_{\ell}$. M\"{u}ller-Gronbach \cite{muller2002optimal} has 
show that for the Milstein scheme \eqref{eq:pclint}  we have 
  \begin{equation} \label{eq:PC}
 \EE \bigl[\sup_{0\le t\le T} \lev x(t)-X^{\ell}(t) \rev^{p}  \bigr]= \bO ( \mid \D t_{\ell} \log(\D t_{\ell}) \mid  ^{p/2}),
\quad p\ge 2,
 \end{equation}
that is the same as for the EM scheme. 
In order to maintain the strong order of convergence we use Brownian Bridge interpolation rather than 
basic piecewise linear interpolation: 
\begin{equation} \label{eq:BBint}
\tX^{\ell}(t) = X^{{\ell}}_n + \lambda_{\ell}\,  (X^{\ell}_{n+1}\!-\!X^{{\ell}}_n) 
+ g(X^{\ell}_{n})\, \left( w(t) - w(n\D t_{\ell}) - \lambda\, \D w^{l}_{n+1} \right),
\end{equation}
for $t\in[n \D t_{\ell}, (n+1) \D t_{\ell} )$. For the Milstein scheme interpolated with Brownian bridges we have 
\cite{muller2002optimal}
\begin{equation*}
 \EE \bigl[\sup_{0\le t\le T} \lev x(t)-\tX^{\ell}(t) \rev^{p}  \bigr] = \bO(  \mid \D t_{\ell} \log(\D t_{\ell}) \mid ^{p}).
 \end{equation*}
Clearly  $\tX^{\ell}(t) $ is not implementable, since in order to construct it, 
the knowledge of the whole trajectory 
$(w(t))_{0\le t \le T}$ is required. However, we will demonstrate that combining $\tX^{\ell}(t)$ with conditional Monte Carlo 
techniques can dramatically improve the convergence of the variance of the MLMC estimator. This is due to the fact that for suitable
MLMC estimators  
only distributional knowledge of certain functionals of  $(w(t))_{0\le t \le T}$ will be required. 

\subsection{Euler-Maruyama scheme}

In this section we demonstrate how to approximate the most common payoffs using the EM scheme \eqref{eq:EM}.

The Asian option we consider has the  payoff
\[
P = \left( T^{-1} \int_0^T x(t) \ dt \!-\! K  \right)^{+}.
\]
Using the piecewise linear interpolation \eqref{eq:pclint} one can obtain the following approximation  
\[
P_{l} \equiv T^{-1} \int_0^T X^{\ell}(t) \ dt = T^{-1} \sum_{n=0}^{2^{\ell}-1} \fracs{1}{2}\, \D t_{\ell} \, 
( X^{\ell}_n \!+\! X^{\ell}_{n+1} ),
\]
% Giles at al. \cite{ghm09}  has 
% proved that in that case $\VV_{\ell}= \bO( \D t_{\ell})$.
Lookback options have payoffs of the form 
\[
P = x(T) - \inf_{0\le t\le T} x(t).
\]
A numerical approximation to this payoff is
\[
P_{\ell} \equiv X^{\ell}_{T} - \inf_{0\le n\le 2^{\ell}} X^{\ell}_{n}.
\]
For both of these payoffs it can be 
proved that $V_{\ell} = \bO(\D t_{\ell})$ \cite{ghm09}.

We now consider a digital option, which pays one unit if the asset at the final time 
exceeds the fixed strike price $K$, and pays zero otherwise.
Thus, the discontinuous payoff function has the form
\[
P = \mathbf{1}_{ \{x(T)>K\} },
\]
with the corresponding EM value
\[
P_{\ell} \equiv \mathbf{1}_{ \{X^{\ell}_{T}>K\} }.
\]
Assuming boundedness of the density of the solution 
to \eqref{eq:SDE} in the neighborhood of the strike $K$, it has been proved
in \cite{ghm09} that $V_{\ell} = \bo(\D t_{\ell}^{1/2-\delta})$, for any $\delta>0$.
This result has been tightened by Avikainen \cite{avikainen09} 
who proved that $V_{\ell} = \bO(\D t_{\ell}^{1/2}\log \D t_{\ell})$.

An up-and-out call gives a European
payoff if the asset never exceeds the barrier, $B$, otherwise it pays zero.
So, for the exact solution we have
\[
P = (x(T)-K)^+ \mathbf{1}_{ \{\sup_{0\le t\le T} x(t) \le B\} },
\]
and for the EM approximation
\[
P_{\ell} \equiv (X^{\ell}_{T}-K)^+ \mathbf{1}_{ \{\sup_{0\le n\le 2^{\ell}} X^{\ell}_{n} \le B\}
}.
\]
A down-and-in call knocks in when the minimum asset
price dips below the barrier $B$, so that
\[
P = (x(T)-K)^+ \mathbf{1}_{ \{\inf_{0\le t\le T} x(t) \le B\} },
\]
and, accordingly,
\[
P_{l} \equiv (X^{\ell}_{T}-K)^+ \mathbf{1}_{ \{\inf_{0\le n\le 2^{\ell}} X^{\ell}_{n} \le B\}
}.
\]
For both of these barrier options we have $\VV_{\ell} = \bo(\D t_{\ell}^{1/2-\delta})$, for any $\delta>0$, assuming that 
 $\inf_{0\le t\le T} x(t)$ and $\sup_{0\le t\le T} x(t)$ have bounded density in the neighborhood 
of $B$ \cite{ghm09}. 

\begin{table}[h!] 
\begin{center} 
\begin{tabular}{|l|l|l|} 
\hline & \multicolumn{2}{c|}{Euler}  \\ 
option & numerical  & analysis   \\ \hline
Lipschitz & $\bO(\D t_{\ell})$       & $\bO(\D t_{\ell})$      \\
Asian     & $\bO(\D t_{\ell})$       & $\bO(\D t_{\ell})$     \\
lookback  & $\bO(\D t_{\ell})$       & $\bO(\D t_{\ell})$       \\
barrier   & $\bO(\D t_{\ell}^{1/2})$ & $\bo(\D t_{\ell}^{1/2-\delta})$  \\
digital   & $\bO(\D t_{\ell}^{1/2})$ & $\bO(\D t_{\ell}^{1/2}\log \D t_{\ell})$  \\ \hline
\end{tabular} 
\end{center}
\caption{Orders of convergence for $V_{_{\ell}}$ as observed numerically and 
proved analytically for both  Euler  discretisations; 
$\delta$ can be any strictly positive constant.} \label{t:E1tab}
\end{table}

As summarised in Table \ref{t:E1tab},
numerical results taken form \cite{giles08b} suggest that all of these results 
are near-optimal.

% \begin{table}[h!]
% \begin{center}
% \begin{tabular}{|l|l|l|l|l|}
% \hline & \multicolumn{2}{c|}{Euler} &  \multicolumn{2}{c|}{Milstein} \\ 
% option & numerical  & analysis & numerical & analysis  \\ \hline
% Lipschitz & $\bO(h)$       & $\bO(h)$              & $\bO(h^2)$ & $\bO(h^2)$ \\
% Asian     & $\bO(h)$       & $\bO(h)$              & $\bO(h^2)$ & $\bO(h^2)$ \\
% lookback  & $\bO(h)$       & $\bO(h)$              & $\bO(h^2)$ & $\bo(h^{2-\delta})$ \\
% barrier   & $\bO(h^{1/2})$ & $\bo(h^{1/2-\delta})$ & $\bO(h^{3/2})$ & $\bo(h^{3/2-\delta})$ \\
% digital   & $\bO(h^{1/2})$ & $\bO(h^{1/2}\log h)$ & $\bO(h^{3/2})$ & $\bo(h^{3/2-\delta})$ \\ \hline
% \end{tabular}
% \end{center}
% 
% \caption{Orders of convergence for $V_l$ as observed numerically and 
% proved analytically for both the Euler and Milstein discretisations; 
% $\delta$ can be any strictly positive constant.}
% 
% \end{table}

\subsection{Milstein scheme} \label{sec:Milstein}

In the scalar case of SDEs \eqref{eq:SDE} (that is with $d=m=1$) the Milstein scheme has the form
\begin{equation} \label{eq:M1D}
X^{\ell}_{n+1} = X^{{\ell}}_{n} + f (X^{2^{l}}_{n}) \D t_{\ell} 
+ g(X^{\ell}_{n})\D w^{\ell}_{n+1} + g^{'}(X^{\ell}_{n})g(X^{\ell}_{n}) (   (\D w^{\ell}_{n+1})^{2} - \D t_{\ell}  ), 
\end{equation}
where $g'\equiv \partial g/\partial x$. 
The analysis of Lipschitz European payoffs and Asian options with Milstein scheme is analogous to EM scheme and it has been proved
in \cite{giles_ancient} that in both these cases $V_{\ell} = \bO(\D t_{\ell}^2)$.

\subsubsection{Lookback options}

For clarity of the exposition 
we will express the fine time-step approximation in terms of the coarse time-step, that is  
$\mathcal{P^{'}}_{\Delta
t_{\ell}}:=\{n\Delta t_{\ell-1} :n=0,\fracs{1}{2},1,1+\fracs{1}{2},2,...,2^{\ell-1}\}$. The partition for the coarse approximation is given by $\mathcal{P}_{\Delta
t_{\ell-1}}:=\{n\Delta t_{\ell-1} :n=0,1,2,...,2^{\ell-1}\}$. Therefore, 
$X^{\ell-1}_{n}$ corresponds to  $X^{\ell}_{n}$ for $n=0,1,2,...,2^{\ell-1}$. 
\newline
For pricing lookback options with the EM scheme, as an approximation of the minimum of the process we have simply taken
$ \min_n X^{\ell}_n $.  This approximation could be improved by taking
\[
X^{\ell}_{min} = 
\min_n \left( X^{\ell}_n - \beta^* g(X^{\ell}_{n}) \D t_{\ell}^{1/2}\right).
%\left( \min_n \hS_n \right) \left(1-\beta^*\sig \sqrt{h}\right).
\]
Here 
$\beta^*\approx 0.5826$ is a constant which corrects the $\bO(\D t_{\ell}^{1/2})$ leading 
order error due to the discrete sampling of the path, and thereby restores $\bO(\D t_{\ell})$ 
weak convergence \cite{bgk97}.  However, using this approximation, the difference 
between the computed minimum values and the fine and coarse paths is $\bO(\D t_{\ell}^{1/2})$,
%\[
%\hS_{min, l} - \hS_{min, {l-1}} = O(h_l^{1/2})
%\]
and hence the variance $V_{\ell}$ is $\bO(\D t_\ell)$, corresponding to $\beta\!=\!1$.
In the previous section, this was acceptable because $\beta\!=\!1$ was the best 
that could be achieved in general with the Euler path discretisation which was 
used, but we now aim to achieve an improved convergence rate using
the Milstein scheme.

In order to improve the convergence, the Brownian Bridge interpolant $\tX^{\ell}(t)$ defined in \eqref{eq:BBint} is used. 
We have
\begin{equation*}
 \begin{split} 
 \min_{0\le t < T} \tX^{\ell}(t)  & =   \min_{0 \le n\le 2^{\ell-1}-\halfs} \, \big[  \min_{n \D t_{l-1} \le t < (n+\halfs)\D t_{l-1} } \tX^{\ell}(t) \big]  \\
&=   \min_{0 \le n\le 2^{\ell-1}-\halfs}   X^{\ell}_{n,min},
\end{split} 
\end{equation*}
where minimum of the fine approximation over the first half of the coarse time-step is given by \cite{glasserman04}
\begin{equation}
X^{\ell}_{n,min} = \fracs{1}{2} \left( X^{\ell}_n + X^{\ell}_{n+\halfs} 
- \sqrt{ \left(X^{\ell}_{n+\halfs} \!-\! X^{\ell}_n\right)^2 - 2\, g(X^{\ell}_{n})^2 \, \D t_{l} \log U^{\ell}_n }\ \right), 
\label{eq:lookback1}
\end{equation}
and minimum of the fine approximation over the second half of the coarse time-step is given by 
\begin{equation}
X^{\ell}_{n+\halfs,min} = \fracs{1}{2} \left( X^{\ell}_{n+\halfs} + X^{\ell}_{n+1} 
- \sqrt{ \left(X^{\ell}_{n+1} \!-\! X^{\ell}_{n+\halfs}\right)^2 - 2\, g(X^{\ell}_{n+\halfs})^2 \,  \D t_{\ell}  \log   
U^{\ell}_{n+\halfs }\ }\right), 
\label{eq:lookback2}
\end{equation}
where $U^{\ell}_n, U^{\ell}_{n+\halfs}$ are uniform random variables on the unit interval. 
% Hence, we take  
% \[
%  P^{f}_{\ell} = \min_{0 \le n\le 2^{\ell}-\halfs} \{ X^{\ell}_{n,min}  \}.
% \]
For the coarse path, in order to improve the MLMC variance a slightly different estimator is used, 
see \eqref{con:MLMC}. 
Using the same Brownian increments as we used on the fine path (to guarantee that we stay on the same path), 
equation \eqref{eq:BBint} is used to define $\tX^{\ell-1}_{n+\halfs} \equiv \tX^{\ell-1}((n+\halfs)\D t_{\ell-1}) $.
%  as
% \[
%  \tX^{\ell-1}_{n+\halfs} = \frac{1}{2} \big[ X^{\ell-1}_{n} + X^{\ell-1}_{n+1} - g(X^{\ell-1}_{n}) 
% \big(  (w(\D t_{\ell-1}) - w((n+1)\D t_{\ell-1}) - \D w^{\ell-1}_{n+1} )   \big)   \big].
% \]
Given this interpolated value, the minimum value over the interval $[n \D t_{\ell-1}, (n+1) \D t_{\ell-1}]$
%(that corresponds to $[[2n \D t_{\ell}, 2(n+1) \D t_{\ell}]$) 
can then be taken to be the 
smaller of the minima for the two intervals $[n \D t_{\ell-1},(n+\halfs) \D t_{\ell-1})$ and $[(n+\halfs) \D t_{\ell-1},(n+1) \D t_{\ell-1})$,
\begin{eqnarray}
X^{\ell-1}_{n,min} &=& \fracs{1}{2} \left( X^{\ell-1}_n + \tX^{\ell-1}_{n+\halfs}    
- \sqrt{ \left(\tX^{\ell-1}_{n+\halfs} \!-\! X^{\ell-1}_n\right)^2 - 2\,(g(X^{\ell-1}_{n}))^2 \, \fracs{\D t_{\ell-1}}{2} \log U^{\ell}_{n} }\ \right),
\nonumber \\
X^{\ell-1}_{n+\halfs,min} &=& \fracs{1}{2} \left( \tX^{\ell-1}_{n+\halfs}  + X^{\ell-1}_{n+1} 
- \sqrt{ \left( X^{\ell-1}_{n+1} \!-\! \tX^{\ell-1}_{n+\halfs}  \right)^2 - 2\, (g(X^{\ell-1}_{n}))^2 \, \fracs{\D t_{\ell-1}}{2} \log U^{\ell}_{n+\halfs)} }\ \right).
\nonumber \\
\label{eq:lookback2}
\end{eqnarray}
Note that $g(X^{\ell-1}_{n})$ is used for both time steps. It is because we used the Brownian Bridge 
with diffusion term $g(X^{\ell-1}_{n})$ to derive both minima.
If we changed  $g(X^{\ell-1}_{n})$ to $g(\tX^{\ell-1}_{n+\halfs})$ in $X^{\ell-1}_{n+\halfs,min}$, this
 would mean that different Brownian Bridges were used on the first
and second half of the coarse time-step and as a consequence condition 
\eqref{con:MLMC} would be violated.
Note also the re-use of the same uniform random numbers $U^{\ell}_{n}$ and $U^{\ell}_{n+\halfs}$ used to compute 
the fine path minimum. The $\min(X^{\ell-1}_{n,min},X^{\ell-1}_{n+\halfs,min})$  has exactly the same distribution as $X^{\ell-1}_{n,min}$, 
since they are both based on the same  Brownian interpolation, and 
therefore equality \eqref{con:MLMC} is satisfied.
% We then have 
% \[
%  P^{c}_{\ell-1} = \min_{0 \le n\le 2^{\ell-1}-1} \{ \min{\{X^{\ell-1}_{n,min}, X^{\ell-1}_{n+\halfs,min} \}}  \}.
% \]
Giles et al. \cite{giles_ancient} proved the following Theorem: 
\begin{theorem}
\label{thm:lookback}
The multilevel approximation for a lookback option which is a uniform Lipschitz function of 
$x(T)$ and $\inf_{[0,T]} x(t)$ has $V_l = \bo(\D t_l^{2-\delta})$ for any $\delta\!>\!0$.
\end{theorem}

\subsection{Conditional Monte Carlo }
Giles \cite{giles08b} and Giles et al. \cite{giles_ancient}  have  shown that combining conditional Monte Carlo 
with MLMC results in superior estimators for various financial payoffs. 

To obtain an improvement in the convergence of the MLMC variance barrier and digital options, 
conditional Monte Carlo methods is employed. We briefly describe it here.
Our goal is to calculate $\EE[P]$. Instead, 
we can write 
\[
 \EE[ P]= \EE\big[ \EE [ P \mid Z  ]\big], 
\]
where $Z$ is a random vector. Hence $\EE [ P \mid Z  ]$ is an unbiased estimator of  $\EE[ P]$. We also have 
\[
 {\rm Var\,}[ P] = \EE\big[ {\rm Var\,}[ P \mid Z  ] \big] + {\rm Var\,} \big[  \EE [ P \mid Z]  \big], 
\]
hence ${\rm Var\,}\big[  \EE [ P \mid Z]  \big]\le {\rm Var\,}( P)$. In the context of MLMC we obtain a better 
variance convergence if we condition on different vectors on the fine and the coarse level. That is 
 on the fine level we take  $\EE [ P^{f} \mid Z^{f}  ]$, where $Z^{f}=\{X^{\ell}_{n}\}_{0\le n \le 2^{\ell}}$.  
On the coarse level instead of taking $\EE [ P^{c} \mid Z^{c}  ]$ with $Z^{c}=\{X^{\ell-1}_{n}\}_{0\le n \le 2^{\ell-1}}$,
we take $\EE [ P^{c} \mid Z^{c}, \tilde{Z}^{c} ]$, where $\tilde{Z}^{c}=\{\tilde{X}^{\ell-1}_{n+\halfs}\}_{0\le n \le 2^{\ell-1}}$
are obtained from equation \eqref{eq:BBint}. 
Condition \eqref{con:MLMC} trivially holds by tower property of conditional expectation 
\[
\EE \left[ \EE [ P^{c} \mid Z^{c}  ]\right] = \EE[ P^c ] = \EE \left[ \EE [ P^{c} \mid Z^{c}, \tilde{Z}^{c} ]\right].
\]

\subsection{Barrier options}

The barrier option which is considered is a down-and-out option for 
which the payoff is a Lipschitz function of the value of the underlying
at maturity, provided the underlying has never dropped below a value $B \in \RR$,
\[
P = f(x(T))\ {\bf 1}_{\{\tau\!>\!T\}}.
\]
The crossing time $\tau$ is defined as 
\[
\tau = \inf_{ t }\left\{ x(t)<B \right\}.
\]
This requires the simulation of $(x(T), \1_{\tau> T}))$.
The simplest method sets
\[
 \tau^{\D t_{\ell}} = \inf_{ n }{\{ X^{\ell}_{n} < B  \}}
\]
and as an approximation takes $( X^{\ell}_{2^{\ell-1}}, \1_{\{ \tau^{\D t_{\ell}} > 2^{\ell-1} \}}  )$. But even if we could
simulate the process $\{x(n \D t_{\ell})\}_{0\le n \le {2^{\ell-1}}}$ it is possible for $\{x(t)\}_{0\le t \le T}$ to cross the barrier
between grid points. Using the Brownian Bridge interpolation we can approximate $\1_{\{\tau> T\}}$ by
\[
 \prod_{n=0}^{{2^{\ell-1}}-\halfs} \1_{\{ X^{\ell}_{n,min} \ge B  \}}.
\]
This suggests following the lookback approximation in computing the 
minimum of both the fine and coarse paths. However, the variance would be larger
in this case because the payoff is a discontinuous function of the minimum.
A better treatment, which is the one used in \cite{giles07b}, 
is to use the conditional Monte Carlo approach to further smooth the payoff. 
Since the process $X^{\ell}_n$ is Markovian we have 
\begin{equation*}
 \begin{split} 
\EE  & \big[ f(X^{\ell}_{2^{\ell-1}}) \prod_{n=0}^{{2^{\ell-1}}-\halfs} \1_{\{ X^{\ell}_{n,min} \ge B  \}}  \big]   \\
   & = \EE \biggl[ \EE \big[ f(X^{\ell}_{2^{\ell-1}}) \prod_{n=0}^{{2^{\ell-1}}-\halfs} \1_{\{ X^{\ell}_{n,min} \ge B  \}}
          \mid X^{\ell}_{0},\ldots,X^{\ell}_{{2^{\ell-1}}} \big] \biggr] \\
    & = \EE \biggl[   f(X^{\ell}_{2^{\ell-1}}) \prod_{n=0}^{{2^{\ell-1}}-\halfs} \EE  \big[\1_{\{ X^{\ell}_{n,min} \ge B  \}} 
        \mid X^{\ell}_{n},X^{\ell}_{n+1} \big] \biggr] \\ 
    & = \EE \biggl[   f(X^{\ell}_{2^{\ell-1}}) \prod_{n=0}^{{2^{\ell-1}}-\halfs} (1-p^{\ell}_{n}) \biggr],
\end{split} 
\end{equation*}
where from \cite{glasserman04} 
\begin{equation*}
 \begin{split}
  p^{\ell}_{n} = & \, \PP\left( \inf_{n \D t_{\ell} \le t <  (n+\halfs)\D t_{\ell}} \tX(t) < B \ | \ X^{\ell}_n, X^{\ell}_{n+\halfs}\right) \\
= & \, \exp\left(\frac{ - 2\, (X^{\ell}_n \!-\! B)^+(X^{\ell}_{n+\halfs} \!-\! B)^+ }{g(X^{\ell}_{n})^2\, \D t_{\ell}} \right),
 \end{split}
\end{equation*}
and 
\begin{equation*}
 \begin{split}
  p^{\ell}_{n+\halfs} = & \, \PP\left( \inf_{(n+\halfs) \D t_{\ell} \le t <  (n+1)\D t_{\ell}} \tX(t) < B \ | \ X^{\ell}_{n+\halfs}, X^{\ell}_{n+1}\right) \\ 
= &  \,\exp\left(\frac{ - 2\, (X^{\ell}_{n+\halfs} \!-\! B)^+(X^{\ell}_{n+1} \!-\! B)^+ }{g(X^{\ell}_{n+\halfs})^2\, \D t_{\ell}} \right).
 \end{split}
\end{equation*}
Hence, for the fine path this gives
\begin{equation} \label{eq:barrier_f}
P^f_{\ell} = f(X^{\ell}_{2^{\ell-1}}) \ \prod_{n=0}^{{2^{\ell-2}}-\halfs} (1 - p^{\ell}_n),
\end{equation}
The payoff for the coarse path is defined similarly. However, in order to reduce the variance, 
we subsample  $\tX^{\ell-1}_{n+\halfs}$, as we did for lookback options, from 
the Brownian Bridge connecting $X^{\ell-1}_{n}$ and $X^{\ell-1}_{n+1}$ 
\begin{equation*}
 \begin{split} 
\EE  & \big[ f(X^{\ell-1}_{2^{\ell-1}}) \prod_{n=0}^{2^{\ell-1}-1} \1_{\{ X^{\ell-1}_{n,min} \ge B  \}}  \big]   \\
   & = \EE \biggl[ \EE \big[ f(X^{\ell-1}_{2^{\ell-1}}) \prod_{n=0}^{2^{\ell-1}-1} \1_{\{ X^{\ell-1}_{n,min} \ge B  \}}
         \mid X^{\ell-1}_{0},\tX^{\ell-1}_{\halfs}, \ldots,\tX^{\ell-1}_{{2^{\ell-1}}-\halfs},X^{\ell-1}_{{2^{\ell-1}}} \big] \biggr] \\
    & = \EE \biggl[   f(X^{\ell-1}_{2^{\ell-1}}) \prod_{n=0}^{{2^{\ell-1}}-1}
 \EE  \big[\1_{\{ X^{\ell-1}_{n,min} \ge B  \}} \mid X^{\ell-1}_{n},\tX^{\ell-1}_{n+\halfs},X^{\ell-1}_{n+1} \big] \biggr] \\ 
    & = \EE \biggl[   f(X^{\ell-1}_{2^{\ell-1}}) \prod_{n=0}^{{2^{\ell-1}}-1} (1-p^{\ell-1}_{1,n})(1-p^{\ell-1}_{2,n}) \biggr],
\end{split} 
\end{equation*}
where 
\[
 p^{\ell-1}_{1,n} = 
\!=\! \exp\left(\frac{ - 2\, (X^{\ell-1}_n \!-\! B)^+(\tX^{\ell-1}_{n+\halfs} \!-\! B)^+ }{g(X^{\ell-1}_{n})^2\, \D t_{\ell}} \right),
\]
and
\[
 p^{\ell-1}_{2,n} = 
\!=\! \exp\left(\frac{ - 2\, (\tX^{\ell-1}_{n+\halfs} \!-\! B)^+(X^{\ell-1}_{n+1} \!-\! B)^+ }{g(X^{\ell-1}_{n})^2\, \D t_{l}} \right).
\]
Note that the same $g(X^{\ell-1}_{n})$ 
is used (rather than using $g(\tX^{\ell-1}_{n+\halfs})$ in $p^{\ell-1}_{2,n}$) to calculate both probabilities for the same reason as we did for lookback options.  
The final estimator can be written as
\begin{equation} \label{eq:barrier_c}
P^c_{\ell-1} = f(X^{\ell-1}_{2^{\ell-1}}) \prod_{n=0}^{{2^{\ell-1}}-1} (1-p^{\ell-1}_{1,n})(1-p^{\ell-1}_{2,n}).
\end{equation}
Giles et al. \cite{giles_ancient} proved the following theorem 
\begin{theorem}
Provided
$\inf_{[0,T]} |g(B)| > 0$,
and 
$\inf_{[0,T]} x(t)$ 
has a bounded density in the neighbourhood of $B$, then
the multilevel estimator for a down-and-out barrier option has variance 
$V_{\ell} = \bo(\D t_{\ell}^{3/2-\delta})$ for any $\delta\!>\!0$.
\end{theorem}

The reason the variance is approximately $\bo(\D t_{\ell}^{3/2-\delta})$ instead of $\bO(\D t_{\ell}^{2})$ 
is the following: due to the strong convergence property
the probability of the numerical approximation being outside $\D t_{\ell}^{1-\delta}$-neighbourhood of the solution to the SDE \eqref{eq:SDE} is arbitrary small, that is for any $\e>0$
\begin{equation} \label{eq:probasympt}
\begin{split}
& \PP \left( \sup_{0\le n \D t_{\ell} \le T} \lev x(n \D t_{\ell}) - X^{\ell}_{n}  \rev \ge \D t_{\ell}^{1-\e}  \right) \\
& \le
 \D t_{\ell}^{-p+p\e} \EE \left[\sup_{0\le n \D t_{\ell} \le T} \lev x(n \D t_{\ell}) - X^{\ell}_{n}  \rev^p \right] =
\bO(\D_{\ell}^{p\e}).
\end{split} 
\end{equation}
If  $\inf_{[0,T]} x(t) $ is outside the $\D t_{\ell}^{1/2}$-neighborhood of the barrier $B$ then by \eqref{eq:probasympt} it is shown that so are numerical approximations. The probabilities of crossing the barrier in that case are asymptotically either $0$ or $1$  and essentially we are in the Lipschitz payoff case. If the $\inf_{[0,T]} x(t) $ is within the $\D t_{\ell}^{1/2}$-neighbourhood of the barrier $B$ then so are the numerical approximations. In that case it can be shown
that $\EE [ ( P^f_{\ell} - P^c_{\ell-1} )^2] = \bO(\D t^{1-\delta})$ but due to the bounded density assumption, the probability that $\inf_{[0,T]} x(t) $ is within  $\D t_{\ell}^{1/2}$-neighbourhood of the barrier $B$ is of order $\D t_{\ell}^{1/2-\delta}$. Therefore the overall MLMC variance is $V_{\ell} = \bo(\D_{\ell}^{3/2-\delta})$ for any $\delta\!>\!0$.

\subsection{Digital options}
\label{subsec:digital}

A digital option has a payoff which is a discontinuous function of the 
value of the underlying asset at maturity, the simplest example being
\[
P = {\bf 1}_{\{x(T) > B\}}.
\]
Approximating $\1_{\{ x(T)> B\}}$ based only on simulations of $x(T)$ by Milstein scheme
will lead to an $\bO(\D t_{\ell})$ fraction of the paths having 
coarse and fine path approximations to $x(T)$ on either side of the strike, 
producing $P_{\ell} - P_{\ell-1} = \pm 1$, resulting in $V_{\ell} = \bO(\D t_{\ell})$.
To improve the variance to $\bO(\D t_{\ell}^{3/2-\delta})$ for all $\delta \!>\!0$,
the conditional Monte Carlo method is used to smooth the payoff (see section 7.2.3 in 
\cite{glasserman04}).
This approach was proved to be successful in Giles et al. \cite{giles_ancient} and was tested numerically in \cite{giles07b},

If $X^{\ell}_{{2^{\ell-1}}-\halfs}$ denotes the value of the fine path approximation 
one time-step before maturity, then the motion thereafter is approximated  
as Brownian motion with constant drift 
$f(X^{\ell}_{{2^{\ell-1}}-\halfs})$ and volatility 
$g(X^{\ell}_{{2^{\ell-1}}-\halfs})$. The conditional 
expectation for the payoff is the probability that $X^{\ell}_{2^{\ell-1}} \!>\! B$ after 
one further time-step, which is
\begin{equation}
P_{\ell}^f = \EE\left[ \1_{\{X^{\ell}_{2^{\ell-1}}>B\}} \mid X^{\ell}_{2^{\ell-1}-\halfs} \right]
 = \Phi \left( \frac{X^{\ell}_{{2^{\ell-1}}-\halfs} \!+\! f(X^{\ell}_{{2^{\ell-1}}-\halfs}) \D t_{\ell} - B}{\mid g(X^{\ell}_{{2^{\ell-1}}-\halfs}) \mid\, \sqrt{\D t_{\ell}}}\right),
\label{eq:digital1}
\end{equation}
where $\Phi$ is the cumulative Normal distribution.

For the coarse path, we note that given the Brownian increment $\D w^{\ell-1}_{2^{\ell-1}-\halfs}$ for the 
first half of the last coarse time-step (which comes from the fine path simulation), 
the probability that $X^{\ell}_{2^{\ell-1}} \!>\! B$ is
\begin{equation} \label{eq:digital2}
\begin{split}
 P_{\ell-1}^c = & \EE\left[ \1_{\{X^{\ell-1}_{2^{\ell-1}}>B\}} \mid X^{\ell-1}_{{2^{\ell-1}}-1} , \D w^{\ell-1}_{2^{\ell-1}-\halfs} \right] \\
= &\Phi \left( \frac{X^{2^{\ell-1}}_{{2^{\ell-1}}-1} \!+\!  f(X^{\ell-1}_{{2^{\ell-1}}-1}) \D t_{\ell-1} 
\!+\! g(X^{\ell-1}_{{2^{\ell-1}}-1})\D w^{\ell-1}_{2^{\ell-1}-\halfs} - B}
{\mid g(X^{2^{\ell-1}}_{{2^{\ell-1}}-1}) \mid\, \sqrt{\D t_{\ell}}}\right).
\end{split}
\end{equation}
The conditional expectation of (\ref{eq:digital2}) is equal to the conditional expectation of 
$P_{\ell-1}^f$ defined by (\ref{eq:digital1}) on level $\ell\!-\!\ell$, and so equality \eqref{con:MLMC}
is satisfied.  
A bound on the variance of the multilevel estimator is given by the following result:
\begin{theorem}
Provided $g(B) \neq 0$, and $x(t)$ has a bounded density in the neighbourhood of $B$, 
then the multilevel estimator for a digital option has variance 
$V_l = \bo(\D t_l^{3/2-\delta})$ for any $\delta\!>\!0$.
\end{theorem}
Results of the above section were tested numerically in \cite{giles08b} and are summarized in the table \ref{t:M}.
\begin{table}[h!] 
\begin{center}
\begin{tabular}{|l|l|l|}
\hline     &  \multicolumn{2}{c|}{Milstein} \\ 
option     & numerical      & analysis  \\ \hline
Lipschitz  & $\bO(\D t_{\ell}^2)$     & $\bO(\D t_{\ell}^2)$ \\
Asian      & $\bO(\D t_{\ell}^2)$     & $\bO(\D t_{\ell}^2)$ \\
lookback   & $\bO(\D t_{\ell}^2)$     & $\bo(\D t_{\ell}^{2-\delta})$ \\
barrier    & $\bO(\D t_{\ell}^{3/2})$ & $\bo(\D t_{\ell}^{3/2-\delta})$ \\
digital    & $\bO(\D t_{\ell}^{3/2})$ & $\bo(\D t_{\ell}^{3/2-\delta})$ \\ \hline
\end{tabular}
\end{center}
\caption{Orders of convergence for $V_l$ as observed numerically and 
proved analytically for Milstein discretisations; 
$\delta$ can be any strictly positive constant.} \label{t:M}
\end{table}

\section{Greeks with MLMC} \label{sec:greeks}

Accurate calculation of prices is only one objective of 
Monte Carlo simulations. Even more important in some ways is the 
calculation of the sensitivities of the prices to various input
parameters.  These sensitivities, known collectively as the ``Greeks'',
are important for risk analysis and mitigation through hedging.

Here we follow the results by Burgos at al. \cite{bg12} to present how MLMC can applied in this setting. 
The pathwise sensitivity approach (also known as Infinitesimal
Perturbation Analysis) is one of the standard techniques
for computing these sensitivities \cite{glasserman04}.
However, the pathwise approach is not applicable when the financial 
payoff function is discontinuous. One solution to these problems is to use the Likelihood Ratio Method 
(LRM) but its weaknesses are that the variance of the resulting estimator
is usually $\bO(\D t_{l}^{-1})$.

Three techniques are presented that improve MLMC variance: 
payoff smoothing using conditional expectations \cite{glasserman04};
an approximation of the above technique using path splitting for the final timestep \cite{ag07};
the use of a hybrid combination of pathwise sensitivity and the Likelihood Ratio Method \cite{giles09}.
We discuss the strengths and weaknesses of these alternatives in different multilevel Monte Carlo settings.

\subsection{Monte Carlo Greeks}

Consider the approximate solution of the general SDE \eqref{eq:SDE}
using  Euler discretisation \eqref{eq:EM}.
The Brownian increments can be defined to be a linear 
transformation of a vector of independent unit Normal random variables $Z$.

The goal is to efficiently estimate the expected value of some financial 
payoff function $P(x(T))$, and numerous first order sensitivities of this
value with respect to different input parameters such as the volatility 
or one component of the initial data $x(0)$. In more general cases $P$ might also depend on the values of
process $\{x(t)\}_{0\le t \le T}$ at intermediate times.

The pathwise sensitivity approach can be viewed as starting
with the expectation expressed as an integral with respect to $Z$:
\begin{equation}
V_{\ell} \equiv \EE\left[ P(X^{\ell}_{n}(Z,\theta)) \right] = \int P(X^{\ell}_{n}(Z,\theta)) \ p_Z(Z) \ d Z.
\label{eq:pathwise_expectation}
\end{equation}
Here $\theta$ represents a generic input parameter, and the probability density 
function for $Z$ is
\[
p_Z(Z) = (2\pi)^{-d/2} \exp\left(-\|Z\|_2^2 / 2\right),
\]
where $d$ is the dimension of the vector $Z$.

Let $X^{\ell}_{n} = X^{\ell}_{n}(Z,\theta)$. If the drift, volatility and payoff functions are all differentiable,
(\ref{eq:pathwise_expectation}) may be differentiated to give
 \begin{equation}
 \frac{\partial V_{\ell}}{\partial \theta} = 
 \int \frac{ \partial P( X^{\ell}_{n} ) }{ \partial X^{\ell}_{n}} \
 \frac{ \partial X^{\ell}_{n}}{\partial \theta} \ p_Z(Z) \ \D Z,
\label{eq:pathwise_sensitivity}
 \end{equation}
with $\displaystyle \frac{\partial X^{\ell}_{n}  }{\partial \theta}$ being obtained by differentiating
\eqref{eq:EM} to obtain
\begin{equation}
\begin{split}
\frac{\partial X^{\ell}_{n+1}}{\partial \theta}  = \frac{\partial X^{\ell}_{n}}{\partial \theta}
&+ \left( \frac{\partial f(X^{\ell}_{n},\o)}{\partial X^{\ell}_n}\, \frac{\partial X^{\ell}_{n}}{\partial \theta}
+ \frac{\partial f(X^{\ell}_{n},\o)}{\partial \o} \right)\D t_{l} \\
&+ \left( \frac{\partial g(X^{\ell}_{n},\o)}{\partial X^{\ell}_{n}}\, \frac{\partial X^{\ell}_{n}}{\partial \theta}
   + \frac{\partial g(X^{\ell}_{n},\o)}{\partial \o}    \right) \Delta w^{l}_{n+1} .
\label{eq:Euler_sensitivity}
\end{split}
\end{equation}
We assume that $Z \rightarrow \Delta w^{l}_{n+1}$ mapping does not depend on $\o$.
It can be proved
that (\ref{eq:pathwise_sensitivity}) remains valid (that is we can interchange integration and differentiation) 
when the payoff function is
continuous and piecewise differentiable, and the numerical estimate obtained by standard Monte Carlo
with $M$ independent path simulations
\[
M^{-1} \sum_{m=1}^M \frac{\partial P(X^{\ell,m}_{n})}{\partial X^{\ell}_{n} }\ \frac{\partial X^{\ell,m}_{n}}{\partial \theta}
\]
is an unbiased estimate for $\partial V / \partial \theta$ with a 
variance which is $\bO(M^{-1})$, if $P(x)$ is Lipschitz and the drift 
and volatility functions satisfy the standard conditions \cite{kp92}.

Performing a change of variables, the expectation can also be expressed as
\begin{equation}
V_{l} \equiv \EE\left[ P(X^{\ell}_{n}) \right] = \int P(x) \ p_{X^{\ell}_{n}}(x,\theta) dx,
\label{eq:LRM_expectation}
\end{equation}
where $p_{X^{\ell}_{n}}(x,\theta)$ is the probability density function for $X^{\ell}_{n}$ which will depend 
on all of the inputs parameters.  Since probability density functions are usually smooth, (\ref{eq:LRM_expectation}) can
be differentiated to give
\[
\frac{\partial V_{\ell}}{\partial \theta}
= \int P(x)\, \frac{\partial p_{X^{\ell}_{n}}}{\partial \theta} \, d x
= \int P(x)\, \frac{\partial (\log p_{X^{\ell}_{n}})}{\partial \theta} \, p_{X^{\ell}_{n}}\  dx
= \EE\left[ P(x)\ \frac{\partial (\log p_{X^{\ell}_{n}})}{\partial \theta} \right].
\]
which can be estimated using the unbiased Monte Carlo estimator 
\[
M^{-1} \sum_{m=1}^M P(X^{\ell,m}_{n})\ \frac{\partial \log p_{X^{\ell}_{n}}(X^{\ell,m}_{n})}{\partial \theta}
\]
This is the Likelihood Ratio Method. Its great advantage is that it 
does not require the differentiation of $P(X^{\ell}_{n})$.  This makes it applicable 
to cases in which the payoff is discontinuous, and it also simplifies the 
practical implementation because banks often have complicated flexible 
procedures through which traders specify payoffs.  However, it does have
a number of limitations, one being a requirement of absolute continuity
which is not satisfied in a few important applications such as the LIBOR 
market model \cite{glasserman04}.

\subsection{Multilevel Monte Carlo Greeks}

The MLMC method for calculating Greeks can be written as
\begin{equation}
\displaystyle \frac{\partial V}{\partial \theta}=\frac{\partial \EE( P)}{\partial \theta}
 \approx \frac{\partial \EE( P_L)}{\partial \theta} 
=\frac{\partial \mathbb{E}( P_0)}{\partial \theta}+\displaystyle\sum_{\ell=1}^{L} 
\frac{\partial \mathbb{E}(P^{f}_{\ell} - P^{c}_{\ell-1})}{\partial \theta}.
\end{equation}
Therefore extending Monte Carlo Greeks to MLMC Greeks is straightforward. However, the challenge
is to keep the MLMC variance small. This can be achieved by appropriate smoothing of the payoff function.   
The techniques that were presented in section \ref{sec:Milstein} are also very useful here. 

\subsection{European call}
\label{burgos:2}
As an example we consider an  European call $P=(x(T)-B)^+$ with $x(t)$ being a geometric Brownian motion with Milstein scheme approximation given by 
\begin{equation} \label{eq:GBEM}
 X^{\ell}_{n+1} = X^{\ell}_{n} + r\,X^{\ell}_{n} \D t_{\ell} + \s \,X^{\ell}_{n}\D w^{\ell}_{n+1} 
+ \frac{\s^2}{2} ( (\D w^{\ell}_{n+1})^2 - \D t_{\ell}  ). 
\end{equation}
We illustrate the techniques by computing delta ($\delta$) and vega ($\nu$), the sensitivities to the asset's initial value $x(0)$ and to its volatility $\sigma$.

Since the payoff is Lipschitz, we can use pathwise sensitivities. We observe that
\begin{equation}  \nonumber 
\frac{\partial }{\partial x} (x-B)^+ = 
\left\{
  \begin{array}{ll}
       0, \ \text{for} \ x<B \\
       1, \ \text{for} \ x>B
    \end{array}
\right.
\end{equation}
This derivative fails to exists when $x=B$, but since this event has probability 0, we may write 
\[
 \frac{\partial }{\partial x} (x-K)^+ = \1_{\{X>B\}}.
\]
Therefore we are essentially dealing with a digital option.

\subsection{Conditional Monte Carlo for Pathwise Sensitivity }

Using conditional expectation the payoff can be smooth as we did it in Section \ref{sec:Milstein}.
European calls can be treated in the exactly the same way as Digital option in Section \ref{sec:Milstein}, that is instead of simulating the whole path, we stop at the penultimate step and then on the last step we consider the full distribution
of $(X^{\ell}_{2^{l}}\mid w^{l}_{0},\ldots,w^{l}_{2^{l}-1})$. 
\newline 
For  digital options  this approach leads 
to \eqref{eq:digital1} and \eqref{eq:digital2}. For the call options we can do analogous calculations.  
In \cite{bg12} numerical results for this approach obtained, with scalar Milstein scheme used to obtain the penultimate step.
They results are presented in Table \ref{t:GCD}.  
For lookback options conditional expectations leads to \eqref{eq:lookback1} and \eqref{eq:lookback2} and for barriers
to \eqref{eq:barrier_f} and \eqref{eq:barrier_c}. Burgos et al \cite{bg12}, applied pathwise sensitivity to these smoothed payoffs, with scalar Milstein scheme used to obtain the penultimate step, and obtained numerical results that we present in Table \ref{t:GLB}.     
\begin{table}[h!] 
\begin{center}
\begin{tabular}{|c|c|c|c|c|}
 \hline   & \multicolumn{2}{c|}{Call}  & \multicolumn{2}{c|}{Digital}      \\  
\hline \textrm{Estimator} & $\beta$ & \textrm{MLMC Complexity} & $\beta$ & \textrm{MLMC Complexity}\\  \hline
\hline \textrm{Value} & $\approx 2.0$ & $\bO(\epsilon^{-2})$& $\approx 1.4$ & $\bO(\epsilon^{-2})$\\ 
\hline \textrm{Delta} & $\approx 1.5$ & $\bO(\epsilon^{-2})$ & $\approx 0.5$ & $\bO(\epsilon^{-2.5})$\\ 
\hline \textrm{Vega} & $\approx 2$ & $\bO(\epsilon^{-2})$ & $\approx 0.6$ & $O(\epsilon^{-2.4})$\\ 
\hline 
\end{tabular} 
\end{center}
 \caption{Orders of convergence for $V_{\ell}$ as observed numerically and 
 corresponding MLMC complexity.} \label{t:GCD}
\end{table}
 
\begin{table}[h!] 
\begin{center}
\begin{tabular}{|c|c|c|c|c|}
 \hline   & \multicolumn{2}{c|}{Lookback}  & \multicolumn{2}{c|}{Barrier}      \\  
\hline \textrm{Estimator} & $\beta$ & \textrm{MLMC Complexity} & $\beta$ & \textrm{MLMC Complexity}\\  \hline
\hline \textrm{Value} & $\approx 1.9$ & $\bO(\epsilon^{-2})$& $\approx 1.6$ & $\bO(\epsilon^{-2})$\\ 
\hline \textrm{Delta} & $\approx 1.9$ & $\bO(\epsilon^{-2})$ & $\approx 0.6$ & $\bO(\epsilon^{-2.4})$\\ 
\hline \textrm{Vega} & $\approx 1.3$ & $\bO(\epsilon^{-2})$ & $\approx 0.6$ & $O(\epsilon^{-2.4})$\\ 
\hline 
\end{tabular} 
\end{center}
 \caption{Orders of convergence for $V_{\ell}$ as observed numerically and 
 corresponding MLMC complexity.}  \label{t:GLB}
\end{table}

\subsection{Split pathwise sensitivities}

There are two difficulties in using conditional 
expectation to smooth payoffs in practice in financial applications. This first is 
that conditional expectation will often 
become a multi-dimensional integral without an obvious closed-form 
value, and the second is that it requires a change to the 
often complex software framework used to specify payoffs. 
As a remedy for these problems the splitting technique to approximate 
$\EE\left[ P(X^{\ell}_{2^{l}}) \mid X^{\ell}_{{2^{\ell}}-1} \right]$ and
$\EE\left[ P(X^{\ell-1}_{2^{\ell-1}}) \mid X^{\ell-1}_{{2^{\ell-1}}-1} , \Delta w^{\ell}_{{2^{\ell}}-2} \right]$,  is used.
We get numerical estimates of these values by ``splitting" every simulated path on the final timestep.
At the fine level: for every simulated path, a set of $s$ final increments
$\{\Delta w^{\ell,i}_{2^{\ell}}\}_{i \in [1,s]}$ is simulated, which can be averaged to get
\begin{equation}
\EE\left[ P(X^{\ell}_{2^{\ell}}) \mid X^{\ell}_{{2^{\ell}}-1} \right] \approx 
\frac{1}{s}\, \sum_{i=1}^{s}P(X^{\ell}_{2^{\ell-1}}, \Delta w^{\ell,i}_{2^{\ell}}   )
\end{equation}
At the coarse level, similar to the case of digital options, the fine increment of the Brownian motion 
over the first half of the coarse timestep is used, 
\begin{equation}
\EE\left[ P(X^{\ell-1}_{2^{\ell-1}}) \mid X^{\ell-1}_{{2^{\ell-1}}-1} , \Delta w^{\ell}_{{2^{\ell}}-2} \right] 
\approx 
\frac{1}{s}\, \sum_{i=1}^{s} P(X^{\ell-1}_{2^{\ell-1}-1}, \Delta w^{\ell}_{{2^{\ell}}-2}, \Delta w^{\ell-1,i}_{{2^{\ell-1}}} )
\end{equation}
This approach was tested in \cite{bg12}, with scalar the Milstein scheme used to obtain the penultimate step, and is presented in Table \ref{t:splitting}. As expected the values of $\beta$  tend to the rates offered by conditional expectations as $s$ increases and the approximation gets more precise.
\begin{table}[h!]
\begin{center}
\begin{tabular}{|c|c|c|c|}
\hline \textrm{Estimator} &$s $& $\beta$ & \textrm{MLMC Complexity}\\ 
\hline \textrm{Value}
					  & $10 $& $\approx 2.0$ & $O(\epsilon^{-2})$\\
					  & $500 $& $\approx 2.0$ & $O(\epsilon^{-2})$\\
\hline \textrm{Delta} 
					  & $10 $& $\approx 1.0$ & $O(\epsilon^{-2}(\log \epsilon)^2)$\\
					  & $500 $& $\approx 1.5$ & $O(\epsilon^{-2})$\\
\hline \textrm{Vega} 
					  & $10 $& $\approx 1.6$ & $O(\epsilon^{-2})$\\
					  & $500 $& $\approx 2.0$ & $O(\epsilon^{-2})$\\
\hline 
\end{tabular} 
\end{center}
 \caption{Orders of convergence for $V_{\ell}$ as observed numerically and 
 the corresponding MLMC complexity.}  \label{t:splitting}
\end{table}

\subsection{Optimal number of samples}

The use of multiple samples to estimate the value of the conditional expectations 
is an example of the splitting technique \cite{ag07}.
If $w$ and $z$ are independent random variables, then for any function $P(w,z)$ 
the estimator
\[
Y_{M,S} = M^{-1} \sum_{m=1}^M \left( S^{-1} \sum_{i=1}^S P(w^{m},z^{(m,i)}) \right)
\]
with independent samples $w^{m}$ and $z^{m,i}$ is an unbiased estimator for
\[
\EE_{w,z} \left[ P(w,z) \right] 
\equiv \EE_w \left[\rule{0in}{0.15in} \EE_z[P(w,z) \,|\, w] \right],
\]
and its variance is 
\[
\VV[Y_{M,S}] = M^{-1}\ \VV_w\left[\rule{0in}{0.15in} \EE_z [P(w,z) \,|\, w] \right]
            + (MS)^{-1}\ \EE_w\left[\rule{0in}{0.15in} \VV_z[P(w,z) \,|\, w] \right].
\]
The cost of computing $Y_{M,S}$ with variance $v_1\, M^{-1} + v_2\, (MS)^{-1}$ is proportional to
\[
c_1\, M + c_2\, M S,
\]
with $c_1$ corresponding to the path calculation and $c_2$ corresponding
to the payoff evaluation. For a fixed computational cost, the variance can 
be minimised by minimising the product
\[
\left( v_1 \!+\! v_2\, s^{-1}\right) \left( c_1 \!+\! c_2\, s\right)
= v_1\, c_2\, s  +  v_1\,c_1 + v_2\, c_2 + v_2\, c_1\, s^{-1},
\]
which gives the optimum value 
$
s_{\mbox{\small opt}} = \sqrt{v_2\, c_1 / v_1\, c_2}
$.
%\[
%s_{\mbox{\small opt}} = \sqrt{\frac{v_2\, c_1}{v_1\, c_2}}.
%\]

%In a standard path calculation 
$c_1$ is $\bO(\D t_{\ell}^{-1})$ since the cost 
is proportional to the number of timesteps, and $c_2$ is $\bO(1)$,
independent of $\D t_{\ell}$.  
If the payoff is Lipschitz, then $v_1$ and $v_2$ are both $\bO(1)$ and
$S_{\mbox{\small opt}}\!=\!\bO(\D t_{\ell}^{-1/2})$.  
% Next, lets consider the payoff with  discontinuity for $x(T) = K$. 
% Noting that under a standard assumptions on SDEs \eqref{eq:SDE} we have
% \[
%  \PP( d(x(T),K) \le \D t^{1/2}  ) = \bO(\D t^{1/2}).
% \]
%  That is $\bO(\D t^{1/2})$ fraction of paths is within $\bO(h^{1/2})$ of the discontinuity. 
% then for these paths
% $\EE_z [g(w,z) \,|\, w] \!=\! \bO(\D t^{-1/2})$ and 
% $\VV_z [g(w,z) \,|\, w] \!=\! \bO(\D t^{-1})$.
% This leads to $v_1$ and $v_2$ both being $\bO(\D t ^{-1/2})$ and so again
% $S_{\mbox{\small opt}}\!=\!\O(\D t^{-1/2})$.
% 
% In both cases, as $\D t\rightarrow 0$, the variance is asymptotically 
% equal to $v_1\, M^{-1}$ and the cost is asymptotically equal to 
% $c_1\, M$.  Thus the use of the vibrato technique does not, to 
% leading order, increase the variance or the computational cost 
% compared to the use of exact conditional expectation in the few 
% cases for which this exists in a simple closed form.

\subsection{Vibrato Monte Carlo}

The idea of vibrato Monte Carlo is to combine pathwise sensitivity and Likelihood Ration Method.  
Adopting the conditional expectation approach,
each path simulation for a particular set of Brownian motion increments 
$w^{\ell} \equiv (\Delta w^{\ell}_1, \Delta w^{\ell}_2, \ldots, \Delta w^{\ell}_{2^{\ell}-1})$ 
(excluding the increment for the final timestep) computes a conditional 
Gaussian probability distribution $p_X(X^{\ell}_{2^{\ell}}|w^{\ell})$. For a 
scalar SDE, if $\mu_{w^{\ell}}$ and $\sigma_{w^{\ell}}$ are the mean and standard deviation 
for given $w^{\ell}$, then
\[
X^{\ell}_{2^l}(w^{\ell},Z) = \mu_{w^{\ell}} + \sigma_{w^{\ell}} Z,
\]
where $Z$ is a unit Normal random variable. The expected payoff can 
then be expressed as
\[
V_{\ell} = \EE \,\left[ \rule{0in}{0.16in} \EE \, [ P(X^{\ell}_{2^{\ell}}) \,|\, w^{{\ell}} ] \right]
= \int \left\{ \int P(x)\ p_{ X^{\ell}_{2^\ell}}(x \,|\, w^{\ell}) \ {\rm d}x \right\} \ p_{w^{\ell}}(y)\, {\rm d }y.
\]
The outer expectation is an average over the discrete Brownian motion
increments, while the inner conditional expectation is 
averaging over $Z$.

To compute the sensitivity to the input parameter $\theta$, the 
first step is to apply the pathwise sensitivity approach for fixed $w^{l}$ 
to obtain
$
\partial \mu_{w^l}/\partial \theta, 
\partial \sigma_{w^l}/\partial \theta.
$
We then apply LRM to the inner conditional expectation to get
\[
\frac{\partial V_{\ell}}{\partial \theta}
\ =\ \EE \,\left[ \frac{\partial }{\partial \theta}\,
 \EE \, \left[ P(X^{\ell}_{2^{\ell}}) \,|\, w^{\ell}\right] \right]
\ =\ \EE \, \left[ \EE_Z\left[ P(X^{\ell}_{2^{\ell}})\ 
\frac{\partial (\log p_{ X^{\ell}_{2^\ell}})}{\partial \theta} \ | \ w^{\ell}\right] \ \right] ,
\]
where 
\[
\frac{\partial (\log p_{ X^{\ell}_{2^\ell}})}{\partial \theta} 
= \frac{\partial (\log p_{ X^{\ell}_{2^\ell}})}{\partial \mu_{w^{\ell}}}\ 
  \frac{\partial \mu_{w^{\ell}}}{\partial \theta}
+ \frac{\partial (\log p_{ X^{\ell}_{2^\ell}})}{\partial \sigma_{w^{\ell}}}\ 
  \frac{\partial \sigma_{w^{\ell}}}{\partial \theta}.
\]
This leads to the estimator
\begin{eqnarray}
\nonumber\displaystyle  \frac{\partial  V_{\ell}}{\partial \theta} \approx
\frac{1}{N_{\ell}} \sum\limits_{m=1}^{N_{\ell}} & ( \displaystyle \frac{\partial \mu_{ w^{\ell,m}}}{\partial \theta}
 \mathbb{E}\,\left[ P \left( X^{\ell}_{2^{\ell}} \right) \frac{\partial (\log p_{{ X^{\ell}_{2^\ell}}})}{\partial \mu_{ w^{\ell}}} | w^{\ell,m} \right] \\ 
 \displaystyle & + \displaystyle \frac{\partial \sigma_{\hat w^{\ell,m}}}{\partial \theta} 
\mathbb{E}\,\left[ P \left( X^{\ell}_{2^{\ell}} \right) \frac{\partial (\log p_{ X^{\ell}_{2^\ell}})}{\partial \sigma_{w^{\ell}}} | w^{\ell,m} \right]  )
 \label{burgos:vmcestimator}
\end{eqnarray}
We compute $\displaystyle \frac{\partial \mu_{ w^{\ell,m}}}{\partial \theta}$ and $\displaystyle \frac{\partial \sigma_{ w^{\ell,m}}}{\partial \theta} $ with pathwise sensitivities. 
With $\displaystyle X^{\ell,m,i}_{2^l}=X^{\ell}_{2^{\ell}}( w^{\ell,m}, Z^{i}) $, we substitute the following estimators into \eqref{burgos:vmcestimator}
\begin{equation} \label{burgos:vmcestimator2}
\left\{\begin{array}{ll}
\displaystyle \mathbb{E}\, \left[ P \left( X^{\ell}_{2^{\ell}} \right) \frac{\partial (\log p_{ X^{\ell}_{2^\ell}})}{\partial \mu_{ w^{\ell}}} | w^{\ell,m} \right] \approx
\frac{1}{s} \sum\limits_{i=1}^s \left( P \left( X^{\ell,m,i}_{2^{\ell}} \right) \frac{X^{2^{\ell},m,i}_{2^{\ell}} - \mu_{ w^{\ell,m}}}{\sigma_{ w^{\ell,m}}^2} \right)
\\
%\displaystyle 
\displaystyle  \mathbb{E}\, \left[ P \left( X^{\ell}_{2^{\ell}}\right) \frac{\partial (\log p_{ X^{\ell}_{2^\ell}})}{\partial \sigma_{ w^{\ell}}} |\hat w^{\ell,m} \right] \approx
\frac{1}{s} \sum\limits_{i=1}^s P \left( X^{\ell,m,i}_{2^{\ell}} \right) \left( -\frac{1}{\sigma_{ w^{\ell,m}}} + \frac{\left( X^{\ell,m,i}_{2^{\ell}} - \mu_{ w^{\ell,m}}\right)^2}{\sigma_{ w^{\ell,m}}^3}\right)
\end{array}
\right.\nonumber
\end{equation}
%which we substitute in \eqref{burgos:vmcestimator} to get the Vibrato Monte Carlo estimator of $\displaystyle \frac{\partial  \hat V}{\partial \theta}$.

In a multilevel setting, at the fine level we can use \eqref{burgos:vmcestimator} directly.
At the coarse level,  as for digital options in section \ref{subsec:digital}, the fine Brownian increments over the first half of the coarse timestep are re-used to derive \eqref{burgos:vmcestimator}.

The numerical experiments for the call option with $s=10$ was obtained \cite{bg12}, with scalar Milstein scheme used to obtain the penultimate step.

\begin{center} 
\begin{tabular}{|c|c|c|}
\hline \textrm{Estimator} & $\beta$ & \textrm{MLMC Complexity}\\ 
\hline \textrm{Value} & $\approx 2.0$ & $O(\epsilon^{-2})$\\ 
\hline \textrm{Delta} & $\approx 1.5$ & $O(\epsilon^{-2})$\\ 
\hline \textrm{Vega} & $\approx 2.0$ & $O(\epsilon^{-2})$\\ 
\hline 
\end{tabular} 
\end{center}

Although the discussion so far has considered an option based on the value 
of a single underlying value at the terminal time $T$, it can be shown that 
the idea extends very naturally to multidimensional cases, producing a 
conditional multivariate Gaussian distribution, and also to financial payoffs 
which are dependent on values at intermediate times. 

\section{MLMC for Jump-diffusion processes}

Giles and Xia in \cite{xg12} investigated the extension of the MLMC method 
%\cite{giles07b,giles08b}
to jump-diffusion SDEs.  We consider models with finite rate activity
% (e.g.~\cite{merton76}), 
using a jump-adapted discretisation in which the jump times are computed
and added to the standard uniform discretisation times.
If the Poisson jump rate is constant, the jump times are the
same on both paths and the multilevel extension is relatively
straightforward, but the implementation is more complex in the
case of state-dependent jump rates for which the jump times
naturally differ.

Merton\cite{merton76} proposed a jump-diffusion process, in which the asset
price follows a jump-diffusion SDE:

\begin{equation}
d x(t)=f(x(t-)) \,\D t + g(x(t-)) \,\D w(t) + c(x(t-))\,\D J(t),\quad
0\leq t\leq T,  \label{eq:jumpSDE}
\end{equation}
where the jump term $J(t)$ is a compound Poisson process 
$\sum_{i=1}^{N(t)}(Y_i-1)$, the jump magnitude ~$Y_i$ has a prescribed
distribution, and $N(t)$ is a Poisson process with intensity $\lambda$,
independent of the Brownian motion. Due to the existence of jumps, the
process is a c\`{a}dl\`{a}g process, i.e. having right continuity with left
limits. We note that $x(t-)$ denotes the left limit of the process while
$x(t)=\lim_{s\rightarrow t+}x(t)$. In \cite{merton76}, Merton also assumed
that $\log Y_i$ has a normal distribution.
% with mean $a$ and variance $b$, namely $\log Y_i \sim N(a,b)$.

% There are several ways in which to generalize the Merton model.  Here we 
% consider one case investigated by Glasserman \& Merener \cite{gm04b}, 
% in which the jump rate depends on the asset price, namely
% $\lambda=\lambda(x(t-))$.

\subsection{A Jump-adapted Milstein discretisation}

To simulate finite activity jump-diffusion processes, Giles and Xia \cite{xg12}
used the jump-adapted approximation from Platen and Bruti-Liberat\cite{pb10}. 
For each path simulation, the set of jump times
$\mathbb{J}=\{\tau_1,\tau_2,\ldots,\tau_m\}$ within the time interval
$[0,T]$ is added to a uniform partition $\mathcal{P}_{\Delta
t_{l}}:=\{n\Delta t_{l} :n=0,1,2,...,2^{l}\}$.
 A combined set of discretisation times is then given by 
$\mathbb{T}=\{0=t_0<t_1<t_2<\ldots<t_M=T\}$ and we define a the length of the timestep as $\D t^{n}_l=t_{n+1}-t_n$. 
Clearly,  $\D t^{n}_l \le \D t_{l}$.

Within each timestep the scalar Milstein discretisation is used 
to approximate the SDE \eqref{eq:jumpSDE}, and then the jump is simulated when the
simulation time is equal to one of the jump times.  This gives the
following numerical method:
\begin{equation}
\begin{aligned} &X^{\ell,-}_{n+1} = X^{\ell}_{n}+ f(X^{\ell}_{n})\,\D t^{n}_{\ell}+g(X^{\ell}_{n})\,\Delta
w^{\ell}_{n+1}+{\textstyle\frac{1}{2}}\,g^{'}(X^{\ell}_{n})\,g(X^{\ell}_{n})\,(\Delta (w^{\ell}_{n})^{2}-\D t_{\ell}^{n}),& \\
&X^{\ell}_{n+1}=\left\{ \begin{aligned} X^{\ell,-}_{n+1}&~
+c(X^{\ell,-}_{n+1})(Y_i-1), &~\space&\mbox{when} ~t_{n+1}=\tau_i;~&\\
X^{\ell,-}_{n+1}&, &~\space&\mbox{otherwise}, \end{aligned}\right. \end{aligned}
\end{equation}
where $X^{\ell,-}_{n}=X^{\ell}_{t_n-}$ is the left limit of the approximated path, 
$\Delta w^{\ell}_{n}$ is the Brownian increment
and $Y_i$ is the jump magnitude at $\tau_i$.

\subsubsection{Multilevel Monte Carlo for constant jump rate}

In the case of the jump-adapted discretisation the telescopic sum \eqref{eq:MLMC} is written down with respect 
to $\D t_{\ell}$ rather than to $\D t^{n}_{\ell}$. 
Therefore, we have to define the 
computational complexity as the expected computational cost since
different paths may have different numbers of jumps.
However, the expected number of jumps is finite and therefore the cost
bound in assumption $iv)$ will still remain valid for an appropriate 
choice of the constant $c_3$.

The MLMC approach for a constant jump rate is
straightforward. The jump times $\tau_j$, which are the same for the 
coarse and fine paths, are simulated by setting
$\tau_j-\tau_{j-1}\sim \exp(\lambda)$. 

Pricing European call and Asian options in this setting is straightforward.
For lookback, barrier and digital options we need to consider 
Brownian bridge interpolations as we did in Section \ref{sec:Milstein}. However, due to 
presence of jumps some small modifications are required. 
To improve convergence we will be looking 
at Brownian bridges between time-steps coming from jump-adapted discretization. 
In order to obtain an interpolated value $\tX^{2^{\ell-1}}_{n+\halfs}$ for the coarse time-step 
a Brownian Bridge interpolation over interval $[k_{n}, \hat{k}_{n}]$ is considered, where
\begin{equation}
 \begin{split}
  k_{n} & = \max{   \{ n \D t^{n}_{\ell-1}, \max{\{\tau\in \mathbb{J}: \tau < (n+\halfs) \D t^{n}_{\ell-1} \}}  \}} \\
\hat{k}_{n} & = \min{   \{ (n +1) \D t^{n}_{\ell-1}, \min{\{\tau\in \mathbb{J}: \tau > (n+\halfs) \D t^{n}_{\ell-1} \}}  \}}.
 \end{split}
\end{equation}
Hence
\begin{equation*} 
\begin{split}
\tX^{\ell-1}_{n+\halfs}  = & X^{\ell-1}_{k_{n}} + \lambda_{\ell-1}\,  (X^{\ell-1}_{\hat{k}_{n}}\!-\!X^{\ell-1}_{k_{n}})  \\
& + g(X^{\ell-1}_{k_{n}})\, \left( w^{\ell}((n+\halfs)\D t_{\ell-1})   - w^{\ell}(k_{n}) - \lambda_{\ell-1}\, (w^{\ell}(\hat{k}_{n}) - w^{\ell}(k_{n}) ) \right)
\end{split}
\end{equation*}
where $\lambda_{\ell-1} \equiv ((n+\halfs)\D t^{n}_{\ell-1} - k_{n})/(\hat{k}_{n} - k_{n})$.

In the same way as in Section \ref{sec:Milstein}, the minima over time-adapted discretization can be derived. 
For the fine time-step we have 
\begin{equation*}
X^{\ell}_{n,min} = \fracs{1}{2} \left( X^{\ell}_n + X^{\ell,-}_{n+\halfs} 
- \sqrt{ \left(X^{\ell,-}_{n+\halfs} \!-\! X^{\ell}_n\right)^2 - 2\, g(X^{\ell}_{n})^2 \, \D t^{n}_{\ell} \log U^{\ell}_n }\ \right). 
\end{equation*}
Notice the use of the left limits $X^{\ell,-}$.
Following discussion in the previous sections, the minima for the coarse time-step can be derived
using interpolated value $\tX^{\ell-1}_{n+\halfs}$.
% \begin{eqnarray*}
% X^{\ell-1}_{n,min} &=& \fracs{1}{2} \Biggl( X^{\ell-1}_n + \tX^{\ell-1}_{n+\halfs}    \\ 
% & -&  \sqrt{ \left(\tX^{\ell-1}_{n+\halfs} \!-\! X^{\ell-1}_n\right)^2 - 2\,(g(X^{\ell-1}_{n}))^2 \, +((n+\halfs)\D t^{n}_{\ell-1} - k_{n}) \log U^{\ell}_{2n} }\ \Biggr),
% \nonumber \\
% X^{\ell-1}_{n+\halfs,min} &=& \fracs{1}{2} \Biggr( \tX^{\ell-1}_{n+\halfs}  + X^{\ell-1}_{n+1} \\
% &- &\sqrt{ \left( X^{\ell-1}_{n+1} \!-\! \tX^{\ell-1}_{n+\halfs}  \right)^2 - 2\, (g(X^{\ell-1}_{n}))^2 \, 
% +(\hat{k}_{n}-(n+\halfs)\D t^{n}_{\ell-1} ) \log U^{\ell}_{2(n+1)} }\ \Biggl).
% \nonumber \\
% \end{eqnarray*}
Deriving the payoffs for lookback and barrier option is now straightforward. 

For digital options, due to  jump-adapted time grid, in order to find conditional expectations,  
we need to look at relations between the last jump time and the
last timestep before expiry. In fact, there are three cases:
\begin{enumerate}
\item The last jump time $\tau$ happens before penultimate fixed-time timestep,
i.e. $\tau<(2^{l-1}-2)\D t_{l}$. 
\item The last jump time is within the last fixed-time timestep , \newline i.e. $\tau >(2^{l-1}-1)\D t_{l}$;
\item The last jump time is within the penultimate fixed-time timestep, \newline i.e.  
$(2^{l-1}-1)\D t_{l}>\tau>(2^{l-1}-2)\D t_{l}$.
\end{enumerate}
With this in mind we can easily write down the payoffs for the coarse and fine approximations as we presented in Section
\ref{subsec:digital}.

% \begin{enumerate}
% \item In case 1, the fine-path grid and coarse-path grid is the same as for continuous diffusion case.
% 
% \item In case 2, we condition both fine and coarse approximation on last jump time $T-\tau$ to obtain 
% \begin{equation}
% P_l^f = \Phi \left( \frac{X^{2^{l}}_{T-\tau} \!+\!
%  f(X^{2^{l}}_{T-\tau}) (T-\tau) - K}{|g(X^{2^{l}}_{T-\tau})\, \sqrt{(T-\tau)}}\right),
% \end{equation}
% \begin{equation}
% P_{l-1}^c = \Phi \left( \frac{X^{2^{l-1}}_{T-\tau} \!+\!
%  f(X^{2^{l-1}}_{T-\tau}) (T-\tau) - K}{|g(X^{2^{l-1}}_{T-\tau})\, \sqrt{(T-\tau)}}\right),
% \end{equation}
% 
% \item In the last case, where $\tau$ denotes the last jump time,
% 
% \begin{equation*}
% P_l^f = \Phi \left( \frac{X^{2^{l}}_{{2^{l}}-1} \!+\! f(X^{2^{l}}_{{2^{l}}-1}) \D t_l - K}{|g(X^{2^{l}}_{{2^{l}}-1})\, \sqrt{\D t_l}}\right),
% \end{equation*}
% 
% \begin{equation*} 
% \begin{split}
%  P_{l-1}^c 
% = &\Phi \left( \frac{X^{2^{l-1}}_{T-\tau} \!+\!  f(X^{2^{l-1}}_{T-\tau}) (T-\tau) 
% \!+\! g(X^{2^{l-1}}_{T-\tau}) (w((2^{l-1}-1)\D t_{l-1}) - w(T-\tau)) -K }
% {|g(X^{2^{l-1}}_{T-\tau})|\, \sqrt{\D t_l}}\right).
% \end{split}
% \end{equation*}
% \end{enumerate}

\subsubsection{MLMC for Path-dependent rates}

In the case of a path-dependent jump rate $\lambda(x(t))$, the
implementation of the multilevel method becomes more difficult because
the coarse and fine path approximations may have jumps at different times.
These differences could lead to a large difference between the coarse and
fine path payoffs, and hence greatly increase the variance of the multilevel
correction. To avoid this, Giles and Xia \cite{xg12} modified the simulation approach of 
Glasserman and Merener \cite{gm04} which uses ``thinning'' to treat 
the case in which $\lambda(x(t),t)$ is bounded. Let us recall the thinning 
property of Poisson processes. Let $(N_{t})_{t\ge0}$ be a Poisson process with intensity
$\l$ and define a new process $Z_t$ by "thinning`` $N_t$: take all the jump times
$(\tau_{n}, n \ge 1)$ corresponding to $N$, keep then with probability $0<p<1$ or delete
then with probability $1-p$, independently from each other.  Now order the jump times that have not been 
deleted: $(\tau^{'}_{n}, n\ge 1)$, and define 
\[
 Z_{t} = \sum_{n\ge 1} \1_{t\ge \tau^{'}_{n}}.
\]
Then the process $Z$ is Poisson process with intensity $p\lambda$.

In our setting,  first a Poisson process with a
constant rate $\lambda_{\sup}$ (which is an upper bound of the 
state-dependent rate) is constructed. This gives a set of candidate jump times, and 
these are then selected as true jump times with probability
$\lambda(x(t),t) / \lambda_{\sup}$.
The following jump-adapted thinning Milstein scheme is obtained
\begin{enumerate}
\item Generate the jump-adapted time grid for a Poisson process
with constant rate $\lambda_{\sup}$;

\item Simulate each timestep using the Milstein discretisation;

\item When the endpoint $t_{n+1}$ is a candidate jump time, 
      generate a uniform random number $U\sim[0,1]$, and if
      $U < p_{t_{n+1} }=\dfrac{\lambda (x(t_{n+1} -),t_{n+1} )}{\lambda _{\sup }}$,
      then accept $t_{n+1}$ as a real jump time and simulate the jump.
\end{enumerate}

In the multilevel implementation, the straightforward application of the above algorithm will result in 
different acceptance probabilities for fine and coarse level. There may 
be some samples in which a jump candidate is accepted for the fine path, 
but not for the coarse path, or vice versa. 
Because of the first order strong convergence, the difference in acceptance
probabilities will be $\bO(\D t_{\ell})$, and hence there is an $\bO(\D t_{\ell})$ probability 
of coarse and fine paths differing in accepting candidate jumps.  Such
differences will give an $\bO(1)$ difference in the payoff value, and
hence the multilevel variance will be $\bO(h)$.  A more detailed analysis 
of this is given in \cite{xia2011multilevel}.

To improve the variance convergence rate,  a change of measure is used so 
that the acceptance probability is the same for both fine and coarse paths.
This is achieved by taking the expectation with respect to a new measure $Q$:
\begin{equation*}
\EE_{P}[P^{f}_{\ell}-P^{c}_{\ell-1}] = 
\EE_{Q}[P^{f}_{\ell}\prod_{\tau }R_{\tau }^{f}-P^{c}_{\ell-1}\prod_{\tau }R_{\tau }^{c}]
\end{equation*}%
where $\tau $ are the jump times. The acceptance probability for a candidate 
jump under the measure $Q$ is defined to be $\frac{1}{2}$ for both coarse 
and fine paths, instead of 
$p_\tau = \lambda (X(\tau -),\tau ) \,/\, \lambda_{\sup }$. 
The corresponding Radon-Nikodym derivatives are
\begin{equation*}
R_{\tau }^{f}=\left\{ \begin{aligned} &2p_{\tau }^{f}, ~&\mbox{if}~
U<\frac{1}{2}&~;\\ &2(1-p_{\tau }^{f}),~&\mbox{if}~ U\geq\frac{1}{2}&~,
\end{aligned}\right. \quad \quad R_{\tau }^{c}=\left\{ \begin{aligned}
&2p_{\tau }^{c}, ~&\mbox{if}~ U<\frac{1}{2}&~;\\ &2(1-p_{\tau
}^{c}),~&\mbox{if}~ U\geq\frac{1}{2}&~, \end{aligned}\right.
\end{equation*}
Since $\VV[ R_{\tau }^{f}-R_{\tau }^{c} ] = \bO(\D t^{2})$
and $\VV [\hP_{\ell}-\hP_{\ell-1}] = \bO(\D t^{2})$, this results in the 
multilevel correction variance
$\VV_Q[\hP_{\ell}\prod_{\tau }R_{\tau }^{f}
    -\hP_{\ell-1}\prod_{\tau }R_{\tau }^{c}]$ being $\bO(\D t^{2})$.

If the analytic formulation is expressed using the same thinning and 
change of measure, the weak error can be decomposed into two
terms as follows:
\[
\EE_Q\left[\hP_{\ell}\prod_{\tau }R_{\tau }^{f} -P\prod_{\tau }R_{\tau }\right]
\ =\ 
\EE_Q\left[ (\hP_{\ell} - P)\ \prod_{\tau }R_{\tau }^{f}\right]\ +\ 
\EE_Q\left[ P\ ( \prod_{\tau }R_{\tau }^{f} - \prod_{\tau }R_{\tau })\right].
\]
Using H{\"o}lder's inequality, the bound 
$\max(R_{\tau },R_{\tau }^f)\leq 2$ and standard results for a 
Poisson process, the first term can be bounded using weak convergence 
results for the constant rate process, and the second term can be 
bounded using the corresponding strong convergence results \cite{xia2011multilevel}. 
This guarantees that the multilevel procedure does converge 
to the correct value.

\subsection{\Levy processes}

Dereich and Heidenreich \cite{dereich2011multilevel}  analysed
approximation methods for both finite and infinite activity 
\Levy driven SDEs with globally Lipschitz payoffs.
They have derived upper bounds for MLMC variance for the class of path dependent payoffs
that are Lipschitz continuous with respect to supremum norm. One of their main findings
is that the rate of MLMC variance converges is closely related to Blumenthal-Getoor index of the driving \Levy process 
that measures the frequency of small jumps.
In \cite{dereich2011multilevel} authors considered SDEs driven by the \Levy process 
\[
 s(t) = \Sigma\, w(t) + L(t) + b\, t, 
\]
where $\Sigma$ is the diffusion coefficient, $L(t)$ is a compensated jump process and $b$ is a drift coefficient.
The simplest treatment is to neglect all the jumps with size smaller than $h$. 
To construct MLMC they took $h^{\ell}$, that is at level $\ell$ they neglected jumps smaller than $h^{\ell}$.
Then similarly as in the previous section, a uniform time discretization $\D t_{\ell}$ 
augmented with jump times is used. Let us denote by $\D L(t)=L(t)-L(t)^{-}$, the jump-discontinuity at time t. The crucial observation is that for $h^{'}>h>0$ the jumps of the process $L^{h^{'}}$ can be obtained 
from those of $L^{h}$ by
\[
 \D L(t)^{h^{'}} = \D L_{t}^{h} \, \1_{\{\mid \D L(t)^{h}\mid> h^{'}\}},
\]
this gives the necessary coupling to obtain a good MLMC variance.  
We define a decreasing and invertible function $g:(0,\infty)\rightarrow (0,\infty)$ such that 
\[
 \int\frac{\mid x \mid^{2}}{h^{2}} \, \wedge \, 1\nu (dx) \le g(h) \quad \text{for all} \quad h>0,
\]
where $\nu$ is a \Levy measure, and for for $\ell\in \NN$ we define
\[
 \D t_{\ell} = 2^{-\ell} \quad \text{and} \quad h^{\ell} = g^{-1}(2^{\ell}). 
\]
With this choice of $\D t_{\ell}$ and $h^{\ell}$, authors in \cite{dereich2011multilevel} analysed the standard Euler-Maruyama scheme 
for \Levy driven SDEs. This approach gives good results for a Blumenthal-Getoor index smaller than one. For a Blumenthal-Getoor
index bigger than one, Gaussian approximation of small jumps gives better results \cite{dereich11}.

\section{Multi-dimensional Milstein scheme} \label{sec:MLMC}

In the previous sections it was shown that by combining a numerical approximation 
with the strong order of convergence $\bO(\D t_{\ell})$ with MLMC results in reduction of 
the computational complexity to estimate expected values of 
functionals of SDE solutions with a root-mean-square error of $\eps$
from $\bO(\eps^{-3})$ to $\bO(\eps^{-2})$. However, in general, to obtain 
a rate of strong convergence higher than  $O(\D t^{1/2})$ requires
simulation, or approximation, of \Levy areas. Giles and Szpruch in \cite{giles2012antithetic} through 
the construction of a suitable 
antithetic multilevel correction estimator, showed that we can avoid the
simulation of \Levy areas and still achieve an $O(\D t^2)$ variance
for smooth payoffs, and almost an $O(\D t^{3/2})$ variance for piecewise  smooth payoffs, 
even though there is only $O(\D t^{1/2})$ strong convergence. 

In the previous sections we have shown that it can be better to use different 
estimators for the finer and coarser of the two levels being 
considered, $P^{f}_{\ell}$ when level $\ell$ is the finer level, 
and $P^{c}_{\ell}$ when level $\ell$ is the coarser level.
In this case, we required that
\begin{equation} 
\EE[P^{f}_{\ell}] = \EE[P^{c}_{\ell}] \quad \hbox{for } \ell=1,\ldots,L,
\end{equation}
so that 
\[
\EE [P^f_{L}] = \EE [P^f_{0}] + \sum_{\ell=1}^{L}\EE [P^{f}_{\ell}-P^{c}_{\ell-1}],
\]
still holds. For lookback, barrier and digital options we showed that we can obtain a 
better MLMC variance by suitable modifying the
estimator on the coarse levels. By further exploiting the flexibility of MLMC, Giles and Szpruch \cite{giles2012antithetic}
modified the estimator
on the fine levels in order to avoid simulation of the \Levy areas.   

\subsection{Antithetic MLMC estimator}

Based on the well-known method of antithetic variates (see for example 
\cite{glasserman04}), the idea for the antithetic estimator is to 
exploit the flexibility of the more general MLMC estimator by defining 
$P^{c}_{\ell-1}$ to be the usual payoff $P(X^c)$ coming from a level 
$\ell\!-\!1$ coarse simulation $X^c$, and defining $P^{f}_{\ell}$ 
to be the average of the payoffs $P(X^{\fp}),P(X^{\fm})$ coming from an 
antithetic pair of level $\ell$ simulations, $X^{\fp}$ and $X^{\fm}$.

$X^{\fp}$ will be defined in a way which corresponds naturally to the 
construction of $X^{c}$.  Its antithetic ``twin'' 
$X^{\fm}$ will be defined so that it has exactly the same distribution 
as  $X^{\fp}$, conditional on $X^c$, which ensures that
$
\EE[ P(X^{\fp}) ] = \EE[P(X^{\fm})]
$
and hence \eqref{con:MLMC} is satisfied, but at the same time
\[
 \left( X^{\fp} - X^c\right)\ \approx\ - \left( X^{\fm} - X^c \right)
\]
and therefore
\[
\left( P(X^{\fp}) - P(X^c)\right)\ \approx\ - \left( P(X^{\fm}) - P(X^c) \right),
\]
so that
$
\fracs{1}{2} \left( P(X^{\fp}) + P(X^{\fm}) \right) \approx P(X^c).
$
This leads to $\fracs{1}{2} \left( P(X^{\fp}) + P(X^{\fm}) \right) - P(X^c)$
having a much smaller variance than the standard estimator $P(X^{f})- P(X^c)$.

We now present a lemma which gives
an upper bound on the convergence of the variance of 
$\fracs{1}{2} \left( P(X^{\fp}) + P(X^{\fm}) \right) - P(X^c)$.

\begin{lemma} \label{lemma:payoff}
If $P\in C^2 (\RR^{d}, \RR)$ and there exist
constants $L_1, L_2$ such that for all $x\in \RR^{d}$ 
\[
\left\| \frac{\partial P}{\partial x} \right\| \le L_1, ~~~
\left\| \frac{\partial^2 P} {\partial x^2} \right\| \le L_2.
\]
then for $p\ge2$,
\begin{eqnarray*}
\lefteqn{\hspace*{-0.1in}
\EE \left[ \left(\fracs{1}{2} (P(X^{\fp}) + P(X^{\fm})) - P(X^c)\right)^p \right] }
\\ &&
\le\ 2^{p-1}\, L_1^p\ 
\EE\left[ \lev \fracs{1}{2}(X^{\fp} \!+\! X^{\fm}) - X^{c} \rev^p \right]
\ +\ 2^{-(p+1)}\, L_2^p\  
\EE\left[ \lev X^{\fp} - X^{\fm} \rev^{2p} \right].
\end{eqnarray*}
\end{lemma}

In the multidimensional SDE applications considered in finance, 
the Milstein approximation with the \Levy areas 
set to zero, combined with the antithetic construction, leads to 
$X^{\fp}-X^{\fm} = O(\D t^{1/2})$ but $ \X^f-X^{c} = O(\D t)$.
Hence, the variance 
$\VV[\fracs{1}{2} (P^{\fp}_{l} \!+\! P^{\fm}_{l}) -P^c_{l-1}]$ 
is $O(\D t^2)$, which is the order obtained for scalar SDEs using 
the Milstein discretisation with its first order strong convergence.

\subsection{Clark-Cameron Example} \label{sec:CC}

The paper of Clark and Cameron \cite{cc80} addresses the question 
of how accurately one can approximate the solution of an SDE driven 
by an underlying multi-dimensional Brownian motion, using only 
uniformly-spaced discrete Brownian increments. Their model problem is
\begin{eqnarray}
{\rm d}x_1(t) &=& {\rm d}w_{1}(t) \nonumber \\
{\rm d}x_2(t) &=& x_1(t)\, {\rm d}w_2(t),
\label{eq:Clark}
\end{eqnarray}
with $x(0)=y(0)=0$, and zero correlation between the two Brownian 
motions $w_1(t)$ and $w_2(t)$.  These equations can be integrated exactly 
over a time interval $[t_n,t_{n+1}]$, where $t_n = n\,\D t$, to give
\begin{eqnarray} \label{eq:CC}
x_1(t_{n+1}) &=& x_1(t_n) + \D w_{1,n} \nonumber \\
x_2(t_{n+1}) &=& x_2(t_n) + x_1(t_n)\D w_{2,n}
 + \fracs{1}{2} \D w_{1,n}\D w_{2,n} +  \fracs{1}{2} A_{12,n}
\end{eqnarray}
where $\D w_{i,n} \equiv w_i(t_{n+1})-w_i(t_n)$, and $A_{12,n}$ is the 
\Levy area defined as
\[
A_{12,n} = \int_{t_n}^{t_{n+1}} 
\left(\rule{0in}{0.13in} w_1(t) \!-\! w_1(t_n) \right) {\rm d} w_2(t) -
\int_{t_n}^{t_{n+1}} \left(\rule{0in}{0.13in} w_2(t) \!-\! w_2(t_n) \right) {\rm d} w_1(t) .
\]
This corresponds exactly to the Milstein discretisation presented 
in (\ref{eq:Milstein}), so for this simple model problem
the Milstein discretisation is exact.

The point of Clark and Cameron's paper is that for {\it any} numerical 
approximation $X(T)$ based solely on the set of discrete Brownian 
increments $\D w$,
\begin{eqnarray*}
\EE[(x_2(T) - X_2(T))^2] 
 &\geq & \fracs{1}{4}\, T\, \D t.
\end{eqnarray*}

Since in this section we use superscript $f,a,c$ for fine $X^{f}$, antithetic $X^{a}$ and coarse $X^{c}$ approximations, respectively, we drop the superscript $\ell$ for the clarity of notation. 

\begin{figure}[b!]
\begin{center}
\includegraphics[width=0.75\textwidth]{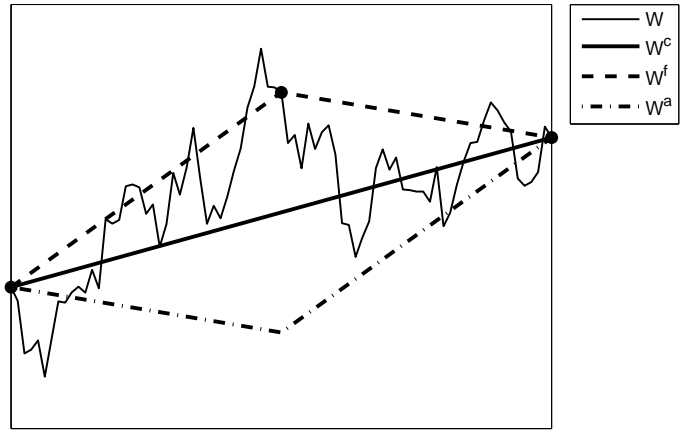}
\end{center}
\vspace{-0.4in}
\caption{Brownian path and approximations over one coarse timestep}
\label{fig:Brownian}
\end{figure}

We define a coarse path approximation $X^c$ with timestep $\D t$
by neglecting the \Levy area terms to give
\begin{eqnarray}
X^c_{1,n+1} &=& X^c_{1,n} + \D w^{\ell-1}_{1,n+1} \nonumber \\
X^c_{2,n+1} &=& X^c_{2,n} + X^c_{1,n}\D w^{\ell-1}_{2,n+1}
 + \fracs{1}{2} \, \D w^{\ell-1}_{1,n+1}\, \D w^{\ell-1}_{2,n+1} \label{eq:CC1}
\end{eqnarray}
This is equivalent to replacing the true Brownian path by a piecewise 
linear approximation as illustrated in Figure \ref{fig:Brownian}.

Similarly, we define the corresponding two half-timesteps of
the first fine path approximation $X^{\fp}$ by
\begin{eqnarray*}
X^{\fp}_{1,n+\halfs} &=& X^{\fp}_{1,n} + \D w^{\ell}_{1,n+\halfs} \nonumber \\
X^{\fp}_{2,n+\halfs} &=& X^{\fp}_{2,n} + X^{\fp}_{1,n}\, \D w^{\ell}_{2,n+\halfs}
 + \fracs{1}{2}\, \D w^{\ell}_{1,n+\halfs}\, \D w^{\ell}_{2,n+\halfs} \nonumber \\
X^{\fp}_{1,n+1} &=& X^{\fp}_{1,n+1} + \D w^{\ell}_{1,n+1}  \\
X^{\fp}_{2,n+1} &=& X^{\fp}_{2,n+\halfs} + X^{\fp}_{1,n+\halfs}\, \D w^{\ell}_{2,n+1}
 + \fracs{1}{2}\, \D w^{\ell}_{1,n+1}\, \D w^{\ell}_{2,n+1} \nonumber
\end{eqnarray*}
where $\D w^{\ell-1}_{n+1} = \D w^{\ell}_{n+\halfs} +  \D w^{\ell}_{n+1}$.
Using this relation, the equations for the two fine timesteps can be 
combined to give an equation for the increment over the coarse timestep,
\begin{eqnarray}
X^{\fp}_{1,n+1} &=& X^{\fp}_{1,n} + \D w^{\ell-1}_{1,n+1} \nonumber \\
X^{\fp}_{2,n+1} &=& X^{\fp}_{2,n} + X^{\fp}_{1,n}\, \D w^{\ell-1}_{2,n+1}
 + \fracs{1}{2} \, \D w^{\ell-1}_{1,n+1}\, \D w^{\ell-1}_{2,n+1} \label{eq:CC2} \\ &&
 +\ \fracs{1}{2} \left(
\D w^{\ell}_{1,n+\halfs}\, \D w^{\ell}_{2,n+1} -
\D w^{\ell}_{2,n+\halfs}\, \D w^{\ell}_{1,n+1}
\right). \nonumber
\end{eqnarray}

The antithetic approximation $X_n^{\fm}$ is defined by exactly 
the same discretisation except that the Brownian increments 
$\delta w_{n}$ and $\delta w_{n+\halfs}$ are swapped, 
as illustrated in Figure \ref{fig:Brownian}.  This gives
\begin{eqnarray*}
X^{\fm}_{1,n+\halfs} &=& X^{\fm}_{1,n} + \D w^{\ell}_{1,n+1}, \nonumber \\
X^{\fm}_{2,n+\halfs} &=& X^{\fm}_{2,n} + X^{\fm}_{1,n}\, \D w^{\ell}_{2,n+1}
 + \fracs{1}{2}\, \D w^{\ell}_{1,n+1}\, \D w^{\ell}_{2,n+1}, \nonumber \\
X^{\fm}_{1,n+1} &=& X^{\fm}_{1,n+\halfs} + \D w^{\ell}_{1,n+\halfs},  \\
X^{\fm}_{2,n+1} &=& X^{\fm}_{2,n+\halfs} + X^{\fm}_{1,n+\halfs}\,\D w^{\ell}_{2,n+\halfs}
 + \fracs{1}{2}\, \D w^{\ell}_{1,n+\halfs}\, \D w^{\ell}_{2,n+\halfs}, \nonumber
\end{eqnarray*}
and hence
\begin{eqnarray}
X^{\fm}_{1,n+1} &=& X^{\fm}_{1,n} + \D w^{\ell-1}_{1,n+1}, \nonumber \\
X^{\fm}_{2,n+1} &=& X^{\fm}_{2,n} + X^{\fm}_{1,n}\, \D w^{\ell-1}_{2,n+1}
 + \fracs{1}{2} \, \D w^{\ell-1}_{1,n+1}\, \D w^{\ell-1}_{2,n+1} \label{eq:CC3} \\ &&
 -\ \fracs{1}{2} \left(
\D w^{\ell}_{1,n+\halfs}\, \D w^{\ell}_{2,n+1} -
\D w^{\ell}_{2,n+\halfs}\, \D w^{\ell}_{1,n+1}
\right).  \nonumber 
\end{eqnarray}
Swapping $\D w^{\ell}_{n+\halfs}$ and $\D w^{\ell}_{n+1}$  does not change the 
distribution of the driving Brownian increments, and hence $X^{\fm}$ 
has exactly the same distribution as $X^{\fp}$.  Note also the 
change in sign in the last term in (\ref{eq:CC2}) compared to 
the corresponding term in (\ref{eq:CC3}). This is important
because these two terms cancel when the two equations are averaged.

These last terms correspond to the \Levy areas for the fine path and
the antithetic path, and the sign reversal is a particular instance of
a more general result for time-reversed Brownian motion, \cite{ks91}.  If 
$(w_{t}, 0\le t \le 1 )$ denotes a Brownian motion
on the time interval $[0,1]$ then the time-reversed Brownian
motion $(z_{t}, 0\le t \le 1 )$ defined by
\begin{equation} \label{eq:TR}
  z_{t} = w_{1} - w_{1-t},
\end{equation}
has exactly the same distribution, and it can be shown that its 
\Levy area is equal in magnitude and opposite in sign to that of $w_t$.
\begin{lemma}
If $X_n^{\fp}$, $X_n^{\fm}$ and $X_n^c$ are as defined above, then
\[
X_{1,n}^{\fp} = X_{1,n}^{\fm} = X_{1,n}^c, ~~~~
\fracs{1}{2} \left( X_{2,n}^{\fp} + X_{2,n}^{\fm} \right) = X_{2,n}^c,
~~~~ \forall n \leq N
\]
and
\[
\EE \left[ \left( X_{2,N}^{\fp} - X_{2,N}^{\fm}\right)^4 \right] 
= \fracs{3}{4}\, T\, (T\!+\!\D t)\, \D t^2.
\]
\end{lemma}
In the next section we will see how this lemma generalizes to non-linear multidimensional SDEs \eqref{eq:SDE}.
\subsection{Milstein discretisation - General theory}
Using the coarse timestep $\D t$, the coarse path approximation $X_n^{c}$, 
is given by the Milstein approximation without the \Levy area term,
\begin{equation*}
 \begin{split}
   X^c_{i,n+1} = &\, X^c_{i,n} + f_i(X^c_n)\,\D t_{\ell-1} 
         + \sum_{j=1}^m  g_{ij}(X^c_n)\,\D w^{\ell-1}_{j,n+1} \\
     & + \sum_{j,k=1}^m h_{ijk}(X^c_n)
\left(\D w_{j,n}\,\D w^{\ell-1}_{k,n+1} - \Omega_{jk}\, \D t_{\ell-1} \right).
 \end{split}
\end{equation*}

The first fine path approximation $X_n^{\fp}$ (that corresponds to $X_n^{c}$) uses the corresponding 
discretisation with timestep $\D t/2$,
\begin{eqnarray} \label{eq:eqn1}
 X^{\fp}_{i,n+\halfs} &=& X^{\fp}_{i,n} + f_i(X^{\fp}_n)\,\D t_{\ell-1}/2 
           + \sum_{j=1}^m g_{ij}(X^{\fp}_n)\,\D w^{\ell}_{j,n+\halfs}
 \\ && 
           +\ \sum_{j,k=1}^m h_{ijk}(X^{\fp}_n)
\left(\D w^{\ell}_{j,n+\halfs}\,\D w^{\ell}_{k,n+\halfs} - \Omega_{jk}\, \D t_{\ell-1}/2 \right),
\nonumber \\
 X^{\fp}_{i,n+1} &=& X^{\fp}_{i,n+\halfs} + f_i(X^{\fp}_{n+\halfs})\,\D t_{\ell-1}/2 
           + \sum_{j=1}^m g_{ij}(X^{\fp}_{n+\halfs})\,\D w^{\ell}_{j,n+1} \label{eq:fine+}
\\ && 
           +\ \sum_{j,k=1}^m h_{ijk}(X^{\fp}_{n+\halfs})
\left(\D w^{\ell}_{j,n+1}\,\D w^{\ell}_{k,n+1} - \Omega_{jk}\, \D t_{\ell-1}/2 \right),
\nonumber
\end{eqnarray}
where 
 $\D w^{\ell-1}_{n+1} = \D w^{\ell}_{n+\halfs} +  \D w^{\ell}_{n+1}$.

The antithetic approximation $X_n^{\fm}$ is defined by exactly the same 
discretisation except that the Brownian increments $D w^{\ell}_{n+\halfs}$ and 
$\D w^{\ell}_{n+1}$ are swapped, so that
\begin{eqnarray} \label{eq:eqn2}
 X^{\fm}_{i,n+\halfs} &=& X^{\fm}_{i,n} + f_i(X^{\fm}_n)\,\D t_{\ell-1}/2 
           + \sum_{j=1}^m g_{ij}(X^{\fm}_n)\,\delta w_{n+\halfs}
\nonumber \\ && 
           +\ \sum_{j,k=1}^m h_{ijk}(X^{\fm}_n)
\left(\D w^{\ell}_{j,n+1}\,\D w^{\ell}_{k,n+1} - \Omega_{jk}\, \D t_{\ell-1}/2 \right),
\nonumber \\
 X^{\fm}_{i,n+1} &=& X^{\fm}_{i,n+\halfs} + f_i(X^{\fm}_{n+\halfs})\,\D t_{\ell-1}/2 
           + \sum_{j=1}^m g_{ij}(X^{\fm}_{n+\halfs})\,\D w^{\ell}_{j,n+\halfs} \label{eq:fine-}
\\ && 
           +\ \sum_{j,k=1}^m h_{ijk}(X^{\fm}_{n+\halfs})
\left(\D w^{\ell}_{j,n+\halfs}\,\D w^{\ell}_{k,n+\halfs} - \Omega_{jk}\, \D t_{\ell-1}/2 \right).
\nonumber
\end{eqnarray}
It can be shown that \cite{giles2012antithetic}
\begin{lemma} \label{lemma:diff}
%Let Assumption \ref{as:gLip} hold. 
For all integers $p\geq 2$, there exists a constant $K_p$ such that
\[
\EE \left[ \max_{0\leq n\leq N} \| X^{\fp}_n - X^{\fm}_n \|^p \right] 
\leq K_p \, \D t^{p/2}.
\]
\end{lemma}
Let's denote the average fine and antithetic path as follows
\[
\X_n^f\equiv\fracs{1}{2}(X_n^{\fp}\!+\!X_n^{\fm}). 
\]
The main results of \cite{giles2012antithetic} is the following theorem:
\begin{theorem} \label{thm:main}
%Let Assumption \ref{as:gLip} hold. 
For all  $p\geq 2$, there exists a constant $K_p$ such that
\[
\EE\left[ \max_{0\leq n\leq N} \| \X_n^f - X_n^c \|^p \right] \leq K_p \, \D t^p. 
\]
\end{theorem}
This together with a classical strong convergence result for Milstein discretization allows 
to estimate the MLMC variance for smooth payoffs. 
n the case of  payoff which is a smooth function of the final 
state $x(T)$, taking
$p\!=\!2$ in Lemma \ref{lemma:payoff}, 
$p\!=\!4$ in Lemma \ref{lemma:diff} and  
$p\!=\!2$ in Theorem \ref{thm:main},
immediately gives the result that the multilevel variance
\[
\VV \left[ \fracs{1}{2} \left( P(X_N^{\fp}) + P(X_N^{\fm}) \right) - P(X_N^c) \right]
\]
has an $O(\D t^2)$ upper bound.  This matches the convergence rate 
for the multilevel method for scalar SDEs using the standard first order
Milstein discretisation, 
and is much better than the $O(\D t)$ convergence obtained with the 
Euler-Maruyama discretisation.

However, very few financial payoff functions are twice differentiable 
on the entire domain $\RR^d$.  A more typical 2D example is a call 
option based on the minimum of two assets, 
\[
P(x(T)) \equiv \max\left(0, \min( x_1(T), x_2(T)) - K \right),
\]
which is piecewise linear, with a discontinuity in the gradient along
the three lines
$(s, K)$, $(K, s)$ and $(s, s)$ for $s \geq K$.
%\begin{eqnarray*}
%(x, K),  && x \geq K, \\
%(K, x),  && x \geq K, \\
%(x, x),  && x \geq K.
%\end{eqnarray*}

To handle such payoffs, an assumption which bounds the 
probability of the solution of the SDE having a value at time $T$ close 
to such lines with discontinuous gradients is needed.
\begin{assp} \label{as:pLip}
The payoff function $P \in C(\RR^d, \RR)$ has a uniform Lipschitz bound,
so that there exists a constant $L$ such that
\[
\left| P(x) - P(y) \right| \leq L \, \left| x - y \right|, 
\quad \forall\, x, y \in \RR^d,
\]
and the first and second derivatives exist, are continuous and have 
uniform bound $L$ at all points $x \not\in K$, where $K$ is a set of 
zero measure, and there exists a constant $c$ such that the probability 
of the SDE solution $x(T)$ being within a neighbourhood of the set $K$ 
has the bound
\[
\PP \left(  
\min_{y\in K} \| x(T) - y \| \leq \Eps
\right) \leq c \ \Eps,
\quad \forall\, \Eps > 0.
\]
\end{assp}
In a 1D context, Assumption \ref{as:pLip} corresponds to an assumption of a locally bounded density for $x(T)$.

Giles and Szpruch in \cite{giles2012antithetic}  proved the following result
\begin{theorem} \label{thm:Lip}
If the payoff satisfies Assumption \ref{as:pLip}, then 
\[
\EE \left[ \left(\fracs{1}{2} (P(X_N^{\fp}) + P(X_N^{\fm})) - P(X_N^c)\right)^2 
\right]
= o(\D t^{3/2-\delta})
\]
for any $\delta >0$.
\end{theorem}

\subsection{Piecewise linear interpolation analysis}

The piecewise linear interpolant $X^{c}(t)$ for the coarse path
is defined within the coarse timestep interval $[t_{k}, t_{k+1}]$ as
\[
X^{c}(t) \equiv  
(1\!-\!\lambda) \, X^{c}_k + \lambda \, X^{c}_{k+1}, \quad
\lambda \equiv \frac{t - t_k}{ t_{k+1}-t_{k}}.
\]
Likewise, the piecewise linear interpolants 
$X^{\fp}(t)$ and $X^{\fm}(t)$ 
are defined on the fine timestep $[t_{k}, t_{k+\halfs}]$ as
\[
X^{\fp}(t) \equiv  
(1\!-\!\lambda) \, X^{\fp}_{k} + \lambda \, X^{\fp}_{k+\halfs}, \quad
X^{\fm}(t) \equiv  
(1\!-\!\lambda) \, X^{\fm}_{k} + \lambda \, X^{\fm}_{k+\halfs}, \quad 
\lambda \equiv \frac{t - t_k}{t_{k+\halfs}-t_{k}},
\]
and there is a corresponding definition for the fine timestep
$[t_{k+\halfs}, t_{k+1}]$. It can be shown that \cite{giles2012antithetic}
\begin{theorem} \label{thm:interp}
%Let Assumption \ref{as:gLip} hold. 
For all  $p\geq 2$, there exists a constant $K_p$ such that
\[
\sup_{0\leq t\leq T} \EE \left[\, \| X^{\fp}(t) - X^{\fm}(t) \|^p \right] 
\leq K_p \, \D t^{p/2},
\]
\[
\sup_{0\leq t\leq T} \EE\left[\, \lev \X^f(t) - X^c(t) \rev^p \right] 
\leq K_p \, \D t^p,
\]
%\[
%\EE\left[ \sup_{0\leq t\leq T} \lev \X^f(t) - X^c(t) \rev^p \right] 
%\leq K_p \, \D t^p\, |\log \D t|^p,
%\]
where $\X^f(t)$ is the average of the piecewise linear
interpolants $X^f(t)$ and $X^a(t)$.
\end{theorem}

For an Asian option, the payoff depends on the average
\[
x_{ave}\ \equiv\ 
T^{-1} \int_0^T x(t) \, {\rm d}t.
\]
This can be approximated by integrating the appropriate piecewise linear
interpolant which gives
\begin{eqnarray*}
X^c_{ave} ~~ \equiv ~~
T^{-1} \int_0^T X^c(t) \, {\rm d}t &=& N^{-1}
\sum_{n=0}^{N-1} \fracs{1}{2} (X^c_{n} + X^c_{n+1}), \\
X^{\fp}_{ave} ~~ \equiv ~~ 
T^{-1} \int_0^T X^{\fp}(t) \, {\rm d}t &=& N^{-1}
\sum_{n=0}^{N-1} \fracs{1}{4}
 (X^{\fp}_{n} + 2 X^{\fp}_{n+\halfs} + X^{\fp}_{n+1}), \\
X^{\fm}_{ave} ~~ \equiv ~~
T^{-1} \int_0^T X^{\fm}(t) \, {\rm d}t &=& N^{-1}
\sum_{n=0}^{N-1} \fracs{1}{4}
 (X^{\fm}_{n} + 2 X^{\fm}_{n+\halfs} + X^{\fm}_{n+1}).
\end{eqnarray*}

Due to H{\"o}lder's inequality,
\begin{equation*}
 \begin{split}
  \EE \left[ \, \lev X^{\fp}_{ave} - X^{\fm}_{ave} \rev^p \right] 
&\ \leq \
T^{-1} \int_0^T 
\EE \left[ \, \lev X^{\fp}(t) - X^{\fm}(t) \rev^p \right]\, {\rm d}t \\
&\ \leq\ 
\sup_{[0,T]} \EE \left[ \, \lev X^{\fp}(t) - X^{\fm}(t) \rev^p \right],
 \end{split}
\end{equation*}

and similarly
\[
\EE \left[ \, \lev \fracs{1}{2}(X^{\fp}_{ave}\!+\!X^{\fm}_{ave})
            - X^c_{ave} \rev^p \right]
\ \leq \ 
\sup_{[0,T]} \EE \left[ \, \lev \X^f(t) - X^c(t) \rev^p \right],
\]
Hence, if the Asian payoff is a smooth function of the average, then 
we obtain a second order bound for the multilevel correction variance.

This analysis can be extended to include payoffs which are a smooth 
function of a number of intermediate variables, each of which is a 
linear functional of the path $x(t)$ of the form
\[
\int_0^T g^T(t)\ x(t)\ \mu({\rm d}t),
\]
for some vector function $g(t)$ and measure $\mu({\rm d}t)$.
This includes weighted averages of $x(t)$ at a number of discrete
times, as well as continuously-weighted averages over the whole
time interval.

As with the European options, the analysis can also be extended to 
payoffs which are Lipschitz functions of the average, and have 
first and second derivatives which exist, and are continuous and
uniformly bounded, except for a set of points $K$ of zero measure.

\begin{assp} \label{as:pLip2}
The payoff $P \in C(\RR^d, \RR)$ has a uniform Lipschitz bound,
so that there exists a constant $L$ such that
\[
\left| P(x) - P(y) \right| \leq L \, \left| x - y \right|, 
\quad \forall\, x, y \in \RR^d,
\]
and the first and second derivatives exist, are continuous and have 
uniform bound $L$ at all points $x \not\in K$, where $K$ is a set of 
zero measure, and there exists a constant $c$ such that the probability 
of $x_{ave}$ being within a neighbourhood of the set $K$ 
has the bound
\[
\PP \left(  
\min_{y\in K} \| x_{ave} - y \| \leq \Eps
\right) \leq c \ \Eps,
\quad \forall\, \Eps > 0.
\]
\end{assp}
\bigskip
\begin{theorem} \label{thm:Lip2}
If the payoff satisfies Assumption \ref{as:pLip2}, then 
\[
\EE \left[ \left(\fracs{1}{2} (P(X_{ave}^{\fp}) + P(X_{ave}^{\fm})) - P(X_{ave}^c)\right)^2 
\right]
= o(\D t^{3/2-\delta})
\]
for any $\delta >0$.
\end{theorem}
We refer the reader to \cite{giles2012antithetic} for more details. 

\subsection{Simulations for antithetic Monte Carlo}

\begin{figure}[ht!]
\begin{center}
\includegraphics[scale=0.75]{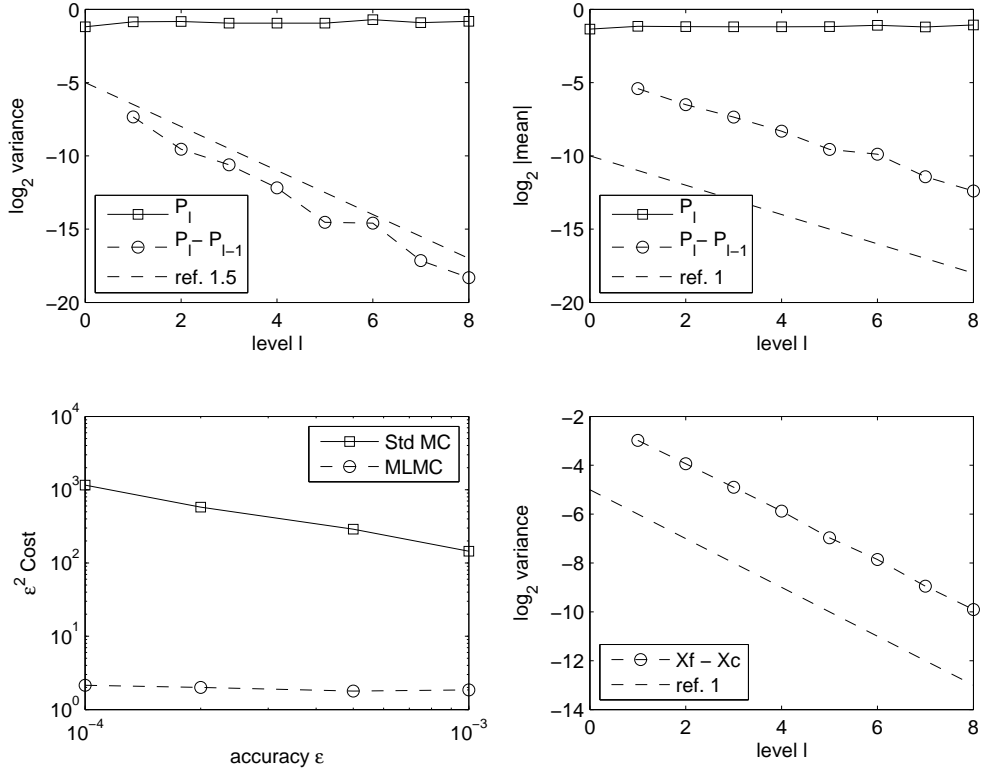}
\end{center}
 \vspace{-20pt}
\caption{Call option.}
\label{fig:Call}
\end{figure}

Here we present numerical simulations for a European option for process simulated by
$X^{f}$, $X^{a}$ and $X^{c}$ defined in section \ref{sec:CC} with initial conditions $x_{1}(0)=x_{2}(0)=1$ . 
The results in Figure \ref{fig:Call} are for a European call option with terminal time 1 and strike 
$K=1$, that is
$
 P = (x(T)-K)^+
$
The top left left plot shows the behaviour of the variance of both $P_{\ell}$ and $P_{\ell}-P_{\ell-1}$. The superimposed reference slope with rate 1.5 indicates that the variance $\VV_{\ell}=\VV[P_{\ell}-P_{\ell-1}] = O (\D t_{\ell}^{1.5})$,
corresponding to $O(\eps^{-2})$ computational complexity of antithetic MLMC. 
The top right plot shows that $\EE[P_{\ell}-P_{\ell-1}]=O(\D t_{\ell})$. 
The bottom left plot shows the computational complexity $C$ (as  defined in Theorem \ref{th:complexity}) with desired accuracy $\eps$. The plot is of $\eps^2\,C$ versus $\eps$, because we expect to see that $\eps^2 \, C$ is only weakly dependent on $\eps$ for MLMC.
For standard Monte Carlo, theory predicts that $\eps^{2}\,C$ should be proportional to the number of timesteps on the finest level, which in turn is roughly proportional to $\eps^{-1}$ due to the weak convergence order. For accuracy $\e=10^{-4}$, 
the antithetic MLMC is approximately  500 times more efficient than the standard Monte Carlo. 
The bottom right plot shows that $\VV[X_{1.\ell}- X_{1.\ell-1}]=O(\D t_{\ell})$. This corresponds to standard strong 
convergence of order 0.5.  
%rate : q$=1.5361$ and residual $=1.1949$
We have also tested the algorithm presented in \cite{giles2012antithetic} for approximation of Asian options. 
Our results were almost identical as for European options. 
In order to treat the lookback, digital and barrier options we found that 
a suitable antithetic approximation to the \Levy areas
are needed. For suitable modification  of the antithetic MLMC estimator
we performed numerical experiments where 
we obtained  $O(\eps^{-2}\log(\eps)^2)$ complexity for estimating barrier, digital and lookback options. 
Currently, we are working on theoretical justification of our results.

\section{Other uses of multilevel method}

\subsection{SPDEs}

Multilevel method has been used for a number of parabolic and elliptic 
SPDE applications \cite{bsz11,cgst11,graubner08} but the first 
use for a financial SPDE is in a new paper by Giles \& Reisinger
\cite{gr12}.

This paper considers an unusual SPDE which results from modelling
credit default probabilities,
\begin{equation}
\label{spde}
\D p = -\mu\, \frac{\partial p}{\partial x}\ \D t 
+ \frac{1}{2} \, \frac{\partial^2  p }{\partial x^2}\ \D t
- \sqrt{\rho}\ \frac{\partial p}{\partial x}\ \D M_t,
~~~ x>0
\end{equation}
subject to boundary condition $p(0,t) \!=\!0$.
Here $p(x,t)$ represents the probability density function for firms 
being a distance $x$ from default at time $t$. The diffusive term 
is due to idiosyncratic factors affecting individual firms, while 
the stochastic term due to the scalar Brownian motion $M_t$ 
corresponds to the systemic movement due to random market effects 
affecting all firms.

Using a Milstein time discretisation with uniform timestep $k$,
and a central space discretisation of the spatial derivatives 
with uniform spacing $h$ gives the numerical approximation
\begin{eqnarray}
p_j^{n+1} &=& p_j^n\ -\ \frac{\mu\, k + \sqrt{\rho\, k}\, Z_n}{2h} 
\left(p_{j+1}^n - p_{j-1}^n\right) 
\nonumber \\&&~~~ +\ \frac{(1\!-\!\rho)\, k + \rho \, k\, Z_n^2}{2h^2} 
 \left(p_{j+1}^n - 2 p_j^n + p_{j-1}^n\right),
\label{discrete}
\end{eqnarray}
where the $Z_n$ are standard Normal random variables so that
$\sqrt{h} Z_n$ corresponds to an increment of the driving scalar
Brownian motion.

The paper shows that the requirment for mean-square stability
as the grid is refined and $k,h\rightarrow 0$ is 
$k/h^2 \leq (1+2\rho^2)^{-1}$, and in addition the accuracy is 
$O(k,h^2)$.  Because of this, the multilevel treatment considers 
a sequence of grids with 
$
h_\ell = 2 \, h_{\ell-1}, ~
k_\ell = 4 \, k_{\ell-1}.
$

The multilevel implementation is very straightforward, with the 
Brownian increments for the fine path being summed pairwise to give
the corresponding Brownian increments for the coarse path.  The 
payoff corresponds to different tranches of a credit derivative
that depends on a numerical approximation of the integral
\[
\int_0^\infty p(x,t) \ \D x.
\]

The computational cost increases by factor 8 on each level, and 
numerical experiments indicate that the variance decreases by 
factor 16.  The MLMC Theorem still applies in this case,
with $\beta=4$ and $\gamma=3$, and so the overall computational 
complexity to achieve an $O(\eps)$ RMS error is again 
$\bO(\eps^{-2} )$.

\subsection{Nested simulation}

The pricing of American options is one of the big challenges for
Monte Carlo methods in computational finance, and
Belomestny \& Schoenmakers have recently written a very interesting
paper on the use of multilevel Monte Carlo for this purpose
\cite{bs11}. Their method is based on Anderson \& Broadie's
dual simulation method \cite{ab04} in which a key component at each 
timestep in the simulation is to estimate a conditional expectation 
using a number of sub-paths.

In their multilevel treatment, Belomestny \& Schoenmakers use the
same uniform timestep on all levels of the simulation.  The quantity 
which changes between different levels of simulation is the number of
sub-samples used to estimate the conditional expectation.

To couple the coarse and fine levels, the fine level uses $N_\ell$
sub-samples, and the coarse level uses $N_{\ell-1} = N_\ell/2$ of 
them. Similar research by N.~Chen
\footnote{unpublished, but presented at the MCQMC12 conference.}
found the multilevel correction variance is reduced if the payoff 
on the coarse level is replaced by an average of the payoffs obtained 
using the first $N_\ell/2$ and the second $N_\ell/2$ samples.
This is similar in some ways to the antithetic approach described in
section \ref{sec:MLMC}.

In future research, Belomestny \& Schoenmakers intend to also 
change the number of timesteps on each level, to increase the 
overall computational benefits of the multilevel approach.

\subsection{Truncated series expansions}

Building on earlier work by Broadie and Kaya \cite{bk06},
Glasserman and Kim have recently developed an efficient 
method \cite{gk11} of exactly simulating the Heston 
stochastic volatility model \cite{heston93}.

The key to their algorithm is a method of representing
the integrated volatility over a time interval $[0,T]$,
conditional on the initial and final values, $v_0$ and $v_T$ as 
\[
\left(\left.\int_0^T V_s \, ds\ \right|\ V_0=v_0, V_T= v_T\right)
\ \stackrel{d}{=}\ 
\sum_{n=1}^\infty x_n + 
\sum_{n=1}^\infty y_n + 
\sum_{n=1}^\infty z_n
\]
where $x_n, y_n, z_n$ are independent random variables.

In practice, they truncate the series expansions at a level which 
ensures the desired accuracy, but a more severe truncation would
lead to a tradeoff between accuracy and computational cost.
This makes the algorithm a candidate for a multilevel treatment
in which level $\ell$ computation performs the truncation at $N_\ell$
(taken to be the same for all three series, for simplicity).

To give more details, the level $\ell$ computation would use
\[
\sum_{n=1}^{N_\ell} x_n + 
\sum_{n=1}^{N_\ell} y_n + 
\sum_{n=1}^{N_\ell} z_n
\]
while the level $\ell \!-\! 1$ computation would use 
\[
\sum_{n=1}^{N_{\ell- 1}} x_n + 
\sum_{n=1}^{N_{\ell- 1}} y_n + 
\sum_{n=1}^{N_{\ell- 1}} z_n
\]
with the same random variables $x_n, y_n, z_n$.

This kind of multilevel treatment has not been tested experimentally, 
but it seems that it might yield some computational savings even though
Glasserman and Kim typically retain only 10 terms in their summations
through the use of a carefully constructed estimator for the truncated 
remainder.  In other circumstances requiring more terms to be retained, 
the savings may be larger.

\subsection{Mixed precision arithmetic}

The final example of the use of multilevel is unusual, because it 
concerns the computer implementation of Monte Carlo algorithms.

In the latest CPUs from Intel and AMD, each core has a vector
unit which can perform 8 single precision or 4 double precision
operations with one instruction.  Together with the obvious fact 
that double precision variables are twice as big as single pecision 
variables and so require twice as much time to transfer, in bulk,
it leads to single precision computations being twice as fast as 
double precision computations.
On GPUs (graphics processors) the difference in performance can be 
even larger, up to a factor of eight in the most extreme cases.

This raises the question of whether single precision arithmetic is
sufficient for Monte Carlo simulation. In general, our view is that
the errors due to single precision arithmetic are much smaller than
the errors due to
\begin{itemize}
\item
statistical error due to Monte Carlo sampling;
\item
bias due to SDE discretisation;
\item
model uncertainty.
\end{itemize}

We have just two concerns with single precision accuracy:
\begin{itemize}
\item
there can be significant errors when averaging the payoffs
unless one uses binary tree summation \cite{higham93} to perform
the summation;

\item
when computing Greeks using ``bumping'', the single precision 
inaccuracy can be greatly amplified if a small bump is used.
\end{itemize}

Our advice would be to always use double precision for the final 
accumulation of payoff values, and pathwise sensitivity analysis
as much as possible for computing Greeks, but if there remains a 
need for the path simulation to be performed in double precision 
then one could use two-level approach in which level 0 corresponds 
to single precision and level 1 corresponds to double precison.

On both levels one would use the same random numbers.  The 
multilevel analysis would then give the optimal allocation of
effort between the single precision and double precision 
computations.  Since it is likely that most of the calculations 
would be single precision, the computational savings would be a 
factor two or more compared to standard double precision calculations.

\section{Multilevel Quasi-Monte Carlo}

In Theorem \ref{th:complexity}, if $\beta>\gamma$, so that rate at which the  
multilevel variance decays with increasing grid level is greater 
than the rate at which the computational cost increases, then the 
dominant computational cost is on the coarsest levels of approximation.

Since coarse levels of approximation correspond to low-dimensional
numerical quadrature, it is quite natural to consider the use of
quasi-Monte Carlo techniques.   This has been investigated 
by Giles \& Waterhouse \cite{gw09} in the context of scalar SDEs
with a Lipschitz payoff.  Using the Milstein approximation with 
a doubling of the number of timesteps on each level gives 
$\beta=2$ and $\gamma=1$.
They used a rank-1 lattice rule to generate the quasi-random 
numbers, randomisation with 32 independent offsets to obtain 
confidence intervals, and a standard Brownian Bridge construction 
of the increments of the driving Brownian process.

\begin{figure}[h!]
\begin{center}
\includegraphics[scale=0.78]{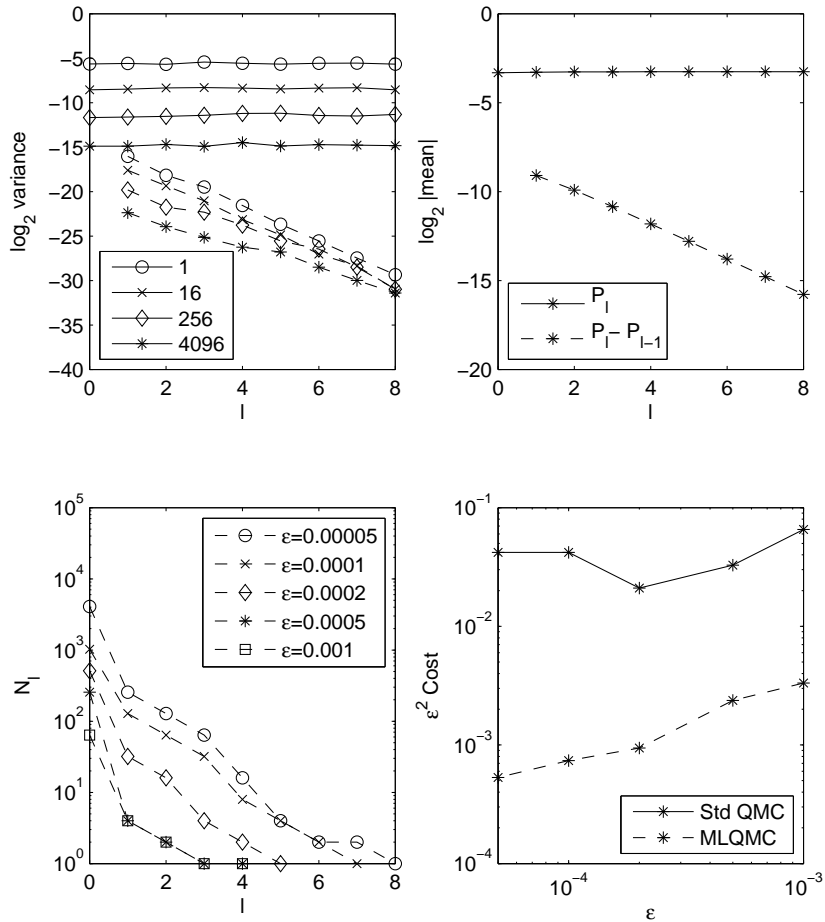}
\end{center}
\vspace{-0.2in}
\caption{European call option (Fig~6.1 from \cite{gw09})}
\label{fig:mlqmc}
\end{figure}

Their empirical observation was that MLMC on its own was better 
than QMC on its own, but the combination of the two was even better.
The QMC treatment greatly reduced the variance per sample for the 
coarsest levels, resulting in significantly reduced costs overall.  
In the simplest case of a European call option, shown in Figure 
\ref{fig:mlqmc}, the top left plot shows the reduction in the variance 
per sample as the number of QMC points is increased.  
The benefit is
much greater on the coarsest levels than on the finest levels.
In the bottom two plots, the number of QMC points on each level is
determined automatically to obtain the required accuracy; 
see \cite{gw09} for the precise details.
Overall, the computational complexity appears to be reduced from 
$O(\eps^{-2})$ to approximately $O(\eps^{-1.5})$.  

Giles \& Waterhouse interpreted the fact that the variance is not 
reduced on the finest levels as being due to a lack of significant 
low-dimensional content. i.e.~the difference in the two payoffs due 
to neighbouring grid levels is due to the difference in resolution 
of the driving Brownian path, 
and this is inherently of high dimensionality.  This suggests that
in other applications with $\beta<\gamma$, which would lead to 
the dominant cost being on the finest levels, then the use of 
quasi-Monte Carlo methods is unlikely to yield any benefits.

Further research is needed in this area to investigate the use 
of other low-discrepancy sequences (e.g.~Sobol) and other ways 
of generating the Brownian increments (e.g.~PCA).
We also refer the reader to \cite{gernol2012} for some results for
randomized multilevel quasi-Monte Carlo.

\section{Conclusions}

In the past 6 years, considerable progress has been achieved
with the multilevel Monte Carlo method for financial options
based on underlying assets described by Brownian diffusions,
jump diffusions, and more general L{\'e}vy processes.

The multilevel approach is conceptually very simple. In essence
it is a recursive control variate strategy, using a coarse path
simulation as a control variate for a fine path simulation,
relying on strong convergence properties to ensure a very
strong correlation between the two.

In practice, the challenge is to couple the coarse and fine path
simulations as tightly as possible, minimising the difference in
the payoffs obtained for each.  In doing this, there is considerable
freedom to be creative, as shown in the use of Brownian Bridge
constructions to improve the variance for lookback and barrier options,
and in the antithetic estimators for multi-dimensinal SDEs which
would require the simulation of \Levy areas to achieve first order
strong convergence.  Another challenge is avoiding large payoff
differences due to discontinuous payoffs; here one can often use
either conditional expectations to smooth the payoff, or a change of
measure to ensure that the coarse and fine paths are on the same side
of the discontinuity.

Overall, multilevel methods are being used for an increasingly
wide range of applications.  This biggest savings are in situations
in which the coarsest approximation is very much cheaper than the finest.
If the finest level of approaximation has only 32 timesteps, then there
are very limited savings to be achieved, but if the finest level has
256 timesteps, then the potential savings are much larger.

Looking to the future, exciting areas for further research include:
\begin{itemize}
\item
more research on multilevel techniques for American and Bermudan
options;

\item
more investigation of multilevel Quasi Monte Carlo methods;

\item
use of multilevel ideas for completely new financial applications,
such as Gaussian copula and new SPDE models.
\end{itemize}

\newpage

\bibliographystyle{plain}

\bibliography{mlmc,mc}
%\bibliography{../../bib/labbrev,../../bib/mlmc,../../bib/mc}

\end{document}